\documentclass[a4paper,fleqn,usenatbib]{mnras}

\usepackage[T1]{fontenc}
\usepackage{ae,aecompl}

\DeclareRobustCommand{\VAN}[3]{#2}
\let\VANthebibliography\thebibliography
\def\thebibliography{\DeclareRobustCommand{\VAN}[3]{##3}\VANthebibliography}

\usepackage{graphicx}	
\usepackage{amsmath}	
\usepackage{amssymb}	

\title[The dual nature of blazar fast variability]{The dual nature of blazar fast variability. Space and ground observations of S5~0716+714}

\author[C. M. Raiteri et al.] 
{C.~M.~Raiteri              $^{ 1}$\thanks{E-mail:claudia.raiteri@inaf.it},
M.~Villata                  $^{ 1}$,
D.~Carosati                 $^{ 2,3}$,
E.~Ben\'{i}tez              $^{ 4}$,
S.~O.~Kurtanidze            $^{ 5,6,7}$,
\newauthor
A.~C.~Gupta                 $^{ 8}$,
D.~O.~Mirzaqulov            $^{ 9}$,
F.~D'Ammando                $^{10}$,
V.~M.~Larionov              $^{11}$,
\newauthor
T.~Pursimo                  $^{12}$,
J.~A.~Acosta-Pulido         $^{13,14}$,
G.~V.~Baida                 $^{15}$,
B.~Balmaverde               $^{ 1}$,
\newauthor
G.~Bonnoli                  $^{16}$\thanks{Recently moved to Instituto de Astrof\'{i}sica de Andaluc\'{i}a (CSIC), Apartado 3004, E-18080 Granada, Spain},
G.~A.~Borman                $^{15}$,
M.~I.~Carnerero             $^{ 1}$,
W.-P.~Chen                  $^{17}$,
V.~Dhiman                   $^{ 8}$,
\newauthor
A.~Di~Maggio                $^{18}$,
S.~A.~Ehgamberdiev          $^{ 9}$,
D.~Hiriart                  $^{19}$,
G.~N.~Kimeridze             $^{ 5}$,
\newauthor
O.~M.~Kurtanidze            $^{ 5,20,7}$,
C.~S.~Lin                   $^{17}$,
J.~M.~Lopez                 $^{21}$,
A.~Marchini                 $^{16}$,
\newauthor
K.~Matsumoto                $^{22}$,
R.~Mujica                   $^{23}$,
M.~Nakamura                 $^{24}$,
A.~A.~Nikiforova            $^{25,11}$,
\newauthor
M.~G.~Nikolashvili          $^{ 5,7}$,
D.~N.~Okhmat                $^{15}$,
J.~Otero-Santos             $^{13,14}$,
N.~Rizzi                    $^{18}$,
\newauthor
T.~Sakamoto                 $^{24}$,
E.~Semkov                   $^{26}$,
L.~A.~Sigua                 $^{ 5}$,
L.~Stiaccini                $^{16}$,
I.~S.~Troitsky              $^{25}$,
\newauthor
A.~L.~Tsai                  $^{17}$,
A.~A.~Vasilyev              $^{25}$,
and A.~V.~Zhovtan            $^{15}$\\
{\it Affiliations are listed at the end of the paper}
}

\date{Accepted XXX. Received YYY; in original form ZZZ}

\pubyear{2020}

\begin{document}
\label{firstpage}
\pagerange{\pageref{firstpage}--\pageref{lastpage}}
\maketitle


\begin{abstract}
Blazar S5~0716+714 is well-known for its short-term variability, down to intra-day time-scales.
We here present the 2-min cadence optical light curve obtained by the {\it TESS} space telescope in 2019 December -- 2020 January and analyse the object fast variability with unprecedented sampling. Supporting observations by the Whole Earth Blazar Telescope Collaboration in $B$, $V$, $R$, and $I$ bands allow us to investigate the spectral variability during the {\it TESS} pointing. The spectral analysis is further extended in frequency to the UV and X-ray bands with data from the {\it  Neil Gehrels Swift Observatory}. 
We develop a new method to unveil the shortest optical variability time-scales. This is based on progressive de-trending of the {\it TESS} light curve by means of cubic spline interpolations through the binned fluxes, with decreasing time bins. The de-trended light curves are then analysed with classical tools for time-series analysis (periodogram, auto-correlation and structure functions). 
The results show that below 3 d there are significant characteristic variability time-scales of about 1.7, 0.5, and 0.2 d. Variability on time-scales $\la 0.2$ d is strongly chromatic and must be ascribed to intrinsic energetic processes involving emitting regions, likely jet sub-structures, with dimension less than about $10^{-3}$ pc.
In contrast, flux changes on time-scales $\ga 0.5$ d are quasi-achromatic and are probably due to Doppler factor changes of geometric origin.

\end{abstract}




\section{Introduction}

Blazars, including BL Lac objects (BL Lacs) and flat-spectrum radio quasars (FSRQ) form a class of active galactic nuclei characterized by extreme and unpredictable emission variability at all frequencies on a wide range of time-scales. In the optical band, small (up to tenths of mag) intra-day flux variations usually overlap on larger (up to several mag) brightness changes on weeks--years scales, likely revealing different physical mechanisms at work. 
The distinctive feature of blazars is that most of the radiation we observe is produced in a relativistic plasma jet pointing toward us. This causes blueshift of the observed radiation with respect to the emitted one, 
contraction of the variability time-scales, 
and flux enhancement by some power of the Doppler factor $\delta=[\Gamma(1-\beta \cos\theta)]^{-1}$, which depends on both the bulk Lorentz factor $\Gamma=(1-\beta^2)^{-1/2}$ (i.e.\ on the bulk velocity $\beta=v/c$ of the emitting plasma) and the viewing angle $\theta$ (i.e.\ the angle between the velocity vector and the line of sight).
Even small changes of $\theta$ can produce large flux variations, and at least part of the blazar variability can likely be due to geometric effects, as in the twisting inhomogeneous jet model proposed by \citet{raiteri2017_nature} to explain the multifrequency long-term behaviour of the FSRQ CTA~102.
In the model by \citet{camenzind1992} instead, rotating plasma bubbles in the jet can produce lighthouse effect and consequently quasi-periodic oscillations (QPOs).
Other variability mechanisms involve intrinsic energetic processes,
like particle injection or acceleration due to shock waves \citep[e.g.][]{blandford1979,marscher1985,sikora2001}, turbulence \citep{marscher2014,pollack2016}, or instabilities that occur in the accretion disc and propagate into the jet, giving rise to QPOs \citep[e.g.][]{king2013}.
Search for periodicities in blazar (and AGN) light curves at all wavelengths have led to a variety of results \citep[e.g.][]{lainela1999,raiteri2001,gierlinski2008,ackermann2015,graham2015,gupta2018,otero2020,penil2020}, and to discussions on their reliability \citep{vaughan2016,covino2019}. Up to now, the $\sim 12 \rm \, yr$ (quasi) periodicity found in the BL Lac object OJ~287 is considered the most robust result and is interpreted in the framework of a binary black hole system \citep[e.g.][but see \citealt{goyal2018}]{sillanpaa1996,villata1998b,valtonen2016}.

Blazars have often shown intra-day/night variability or even microvariability\footnote{In the paper we will use these definitions with some flexibility to indicate flux changes on time-scales shorter than 24 h (intra-day variability) or much shorter than this (microvariability); the term ``intra-night" is used when mentioning ground-based optical observations.} at all wavelengths \citep{miller1989,wagner1995,aharonian2007,albert2007}. This is mainly observed during flaring states, which may be an observing bias, due to the fact that intensive monitoring is usually performed when the source is bright. Moreover, data are more precise when the flux is higher, which is a requirement to assess the smallest and fastest variations. But it may also be the consequence of the amplification of the variability amplitude and contraction of its time-scales due to Doppler enhancement \citep{raiteri2017_nature}.

The {\it Kepler} space telescope, which was in operation between 2009 and 2013, and the following {\it K2} mission in 2014--2018, allowed astrophysicists to investigate the variability of a few blazars on minute time-scales over time spans of several weeks.
\citet{edelson2013} analysed the 181-d long {\it Kepler} light curve of the BL Lac object W2R~1926+42 with a 30-min sampling, showing significant variations on hour scale and many day-scale flares. They found that the power spectral density (PSD) cannot be satisfactorily fitted by simple mathematical laws, implying that blazar variability requires complex physical pictures.
From the study of the long-term {\it Kepler} light curve of the same object,
\citet{mohan2016} inferred a black hole mass of $(1.5-5.9) \times 10^7 \, M_\odot$ and a distance $\leq 1.75 \rm \, pc$ of the emitting region from the jet base. They also found a QPO with 9.1-d period that they ascribed to jet-based processes.
The 100-d long {\it Kepler} light curve with 1-min sampling of W2R~1926+42
was analysed by \citet{sasada2017}. They noticed variations on both intra-day and inter-day scales with a variety of profiles that in general are not symmetric, showing longer decaying than rising times. From this they concluded that microvariability is likely due to intrinsic rather than geometric processes.
An analysis of the multiwavelength behaviour of OJ~287 on multiple time-scales 
was carried out by \citet{goyal2018}, including {\it Kepler} short-cadence data. 
They found a complex shape of the PSD and no QPOs.
{\it Kepler} data of several AGN were investigated by \citet{wehrle2013} and \citet{wehrle2019}. The slopes of the PSD suggested either turbulence behind a shock or instabilities in the accretion disc as the origin of the observed variability.

After {\it Kepler}, the {\it Transiting Exoplanet Survey Satellite} \citep[{\it TESS};][]{ricker2015} was launched in 2018 April with the main aim of performing an all-sky survey to discover transiting exoplanets. However, a small amount of observing time was available for other scientific objectives.
{\it TESS} data include 30-min cadence full frame images and 2-min cadence light curves for selected objects.
In the first year of operation {\it TESS} monitored the South ecliptic sky. 
In the second year the observations shifted to the northern sky. Through the Guest Observer Program, we asked for 2-min cadence observations of a small sample of optically and $\gamma$-ray bright blazars to investigate their microvariability.
The BL Lac object S5~0716+714 was among these sources.
To check the {\it TESS} calibration and to obtain information on the source spectral variability, we also activated the 
Whole Earth Blazar Telescope\footnote{http://www.oato.inaf.it/blazars/webt} (WEBT) Collaboration and required target-of-opportunity (ToO) observations by the {\it Neil Gehrels Swift Observatory} satellite during the {\it TESS} pointing. 

S5~0716+714 at redshift $z=0.31\pm 0.08$ \citep{nilsson2008} is well known for its intra-day variability \citep[e.g.][and references therein]{wagner1996,villata2000,sasada2008,chandra2011,kaur2018}.
An intensive 5-d campaign was organized by the WEBT in March 2014. The source showed microvariability with about 0.1 mag amplitude on hour time-scale, superimposed on wider variations on day scales. The fast events were characterized by high polarization degree \citep{bhatta2016}.
Simultaneous flux changes in the optical and radio bands and in the optical and $\gamma$-ray bands were detected by \citet{villata2008} and \citet{chen2008} during an active state in 2007, which was monitored by the WEBT at low frequencies and by the {\it Astrorivelatore Gamma ad Immagini LEggero} ({\it AGILE}) satellite at $\gamma$ rays.
Search for periodicities in both the optical and $\gamma$-ray light curves by \citet{sandrinelli2017} led to negative results.
In contrast, \citet{rani2013} found evidence for a 63-d period in the optical and \citet{prokhorov2017} identified a 346-d period in $\gamma$ rays. QPOs with time-scales in the range 25--73 min and of $\sim 15$ min were found by \citet{gupta2009} and \citet{rani2010}, respectively, in selected intra-night optical observations.
The multifrequency and polarimetric behaviour of the source during the 2011 October outburst was explained by \citet{larionov2013} in the framework of a shock wave travelling downstream the jet along a helical path.
RadioAstron Space VLBI observations revealed a complex structure of the radio core and inner jet bending \citep{kravchenko2020}.

We here present the 2-min cadence light curve of S5~0716+714 obtained by {\it TESS} from 2019 December 24 to 2020 January 21, complemented by multiband optical data acquired by the WEBT and by ultraviolet and X-ray data by {\it Swift}.
The paper is organized as follows. Section~\ref{sectess} presents the {\it TESS} light curve and Section~\ref{secwebt} the WEBT optical observations, which are used 
to investigate the optical spectral variability. 
The ultraviolet and X-ray data obtained by {\it Swift} are analysed in Section~\ref{secx}.
Section~\ref{sectsa} presents the time series analysis on the {\it TESS} fluxes. The chromatism of the flux variations is discussed in Section~\ref{seccro}. Final considerations and conclusions are drawn in Section~\ref{seclast}.

\section{Observations by {\it TESS}} 
\label{sectess}
\begin{figure}
\includegraphics[width=\columnwidth]{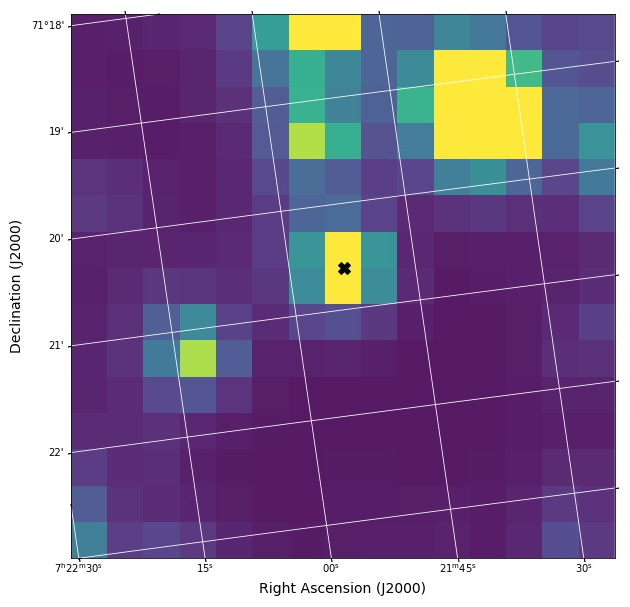}    
\caption{A $15 \times 15$ pixel (1 pixel = 21 arcsec) image of the {\it TESS} field centred on S5~0716+714 (black cross). }
    \label{tess_field}
\end{figure}

Fig.~\ref{tess_field} shows a $15 \times 15$ pixel (1 pixel = 21 arcsec) image of the {\it TESS} field centred on S5~0716+714.

\begin{figure}
	\includegraphics[width=\columnwidth]{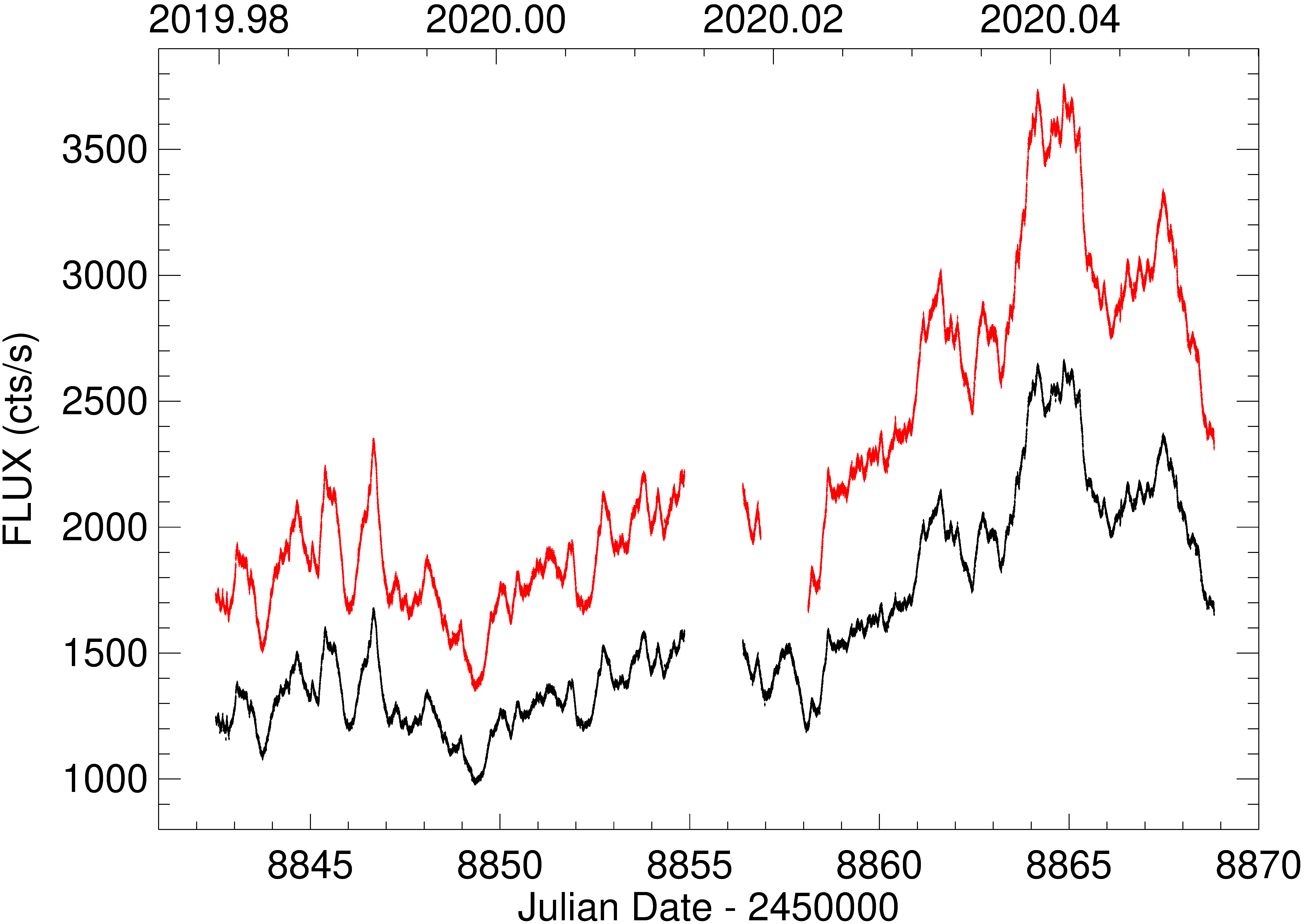}
    \caption{{\it TESS} light curves of S5~0716+714. The lower (black) curve represents SAP\textunderscore FLUX, while the upper (red) curve represents PDCSAP\textunderscore FLUX. }
    \label{tess}
\end{figure}

The 2-min cadence {\it TESS} light curves available at the Mikulski Archive for Space Telescopes\footnote{https://mast.stsci.edu/portal/Mashup/Clients/Mast/\\Portal.html} (MAST) contain both the simple aperture photometry flux (SAP\textunderscore FLUX, electrons per second) and the pre-search data conditioned simple aperture photometry flux (PDCSAP\textunderscore FLUX), which underwent additional instrumental de-trending. 
Both light curves for S5~0716+714 are shown in Fig.~\ref{tess}. 
Apart from a flux shift that has no influence on our variability analysis, they share the same behaviour. The most remarkable difference is that while the SAP\textunderscore FLUX shows only one $\sim 1.5$-d gap at $\rm JD=2458854.9$--8856.4, the PDCSAP\textunderscore FLUX presents also another gap at about 2458857. This can be explained by the fact that de-trending has been optimized for the search for exoplanets, and it is known that this can remove some of the astrophysical signals\footnote{https://outerspace.stsci.edu/display/TESS/2.1+Levels+of+\\data+processing}. 
To verify the stability of the {\it TESS} photometry, we also extracted 30-min light curves from the Full Frame Images (FFI) with the {\small eleanor} open source python package \citep{feinstein2019}.
Time-stacked postcards ($148 \times 104$ pixel cutout regions of the FFIs) are created and background subtracted. Different photometric apertures (2, 3, and 4 pixels) are tested for an optimal light curve extraction. After the raw photometry for each aperture is done,  {\small eleanor} 
performs principal component analysis (PCA) removing all the instrument systematics to produce a corrected light curve. However, different extraction choices led to quite different light curves. 
Therefore, we decided to use the MAST SAP\textunderscore FLUX for our variability analysis, checking its reliability through a comparison with the ground-based WEBT data (Section~\ref{secwebt}).
Errors on SAP fluxes are extremely small, ranging from 0.3 to 0.6 per cent.

The total  number of cadences is $18954$, while the number of cadences with valid (discarding null values and one strong outlier) SAP\textunderscore FLUX is 17812. On the time span of $\sim 26.3 \rm \, d$, the source flux varied by a factor of about 2.7, with a fractional root mean square variability amplitude \citep[e.g.][]{vaughan2003} of $\sim 25$ per cent.
The light curve behaviour looks like produced by a random walk (red) noise process, with small-amplitude and fast flux changes superposed on a smoother, long-term and large-amplitude variability.

\section{Observations by the WEBT}
\label{secwebt}
S5~0716+714 is one of the blazars regularly monitored by the WEBT Collaboration and belongs to the target list of the GLAST-AGILE Support Program \citep[GASP;][]{villata2008,villata2009b}, which started in 2007.
Previous specific works on this source that involved the WEBT were published in \citet{villata2000,ostorero2006,villata2008,chen2008} and \citet{bhatta2016}.

\begin{table}
	\centering
	\caption{WEBT Observatories supporting the {\it TESS} observations of S5~0716+714. The telescope size  is reported (cm), together with the symbol and colour used in the light curve plots.}
	\label{list}
	\begin{tabular}{llrcl}
		\hline
		Observatory & Country &  Tel.~size & Symbol & Colour\\
                            &         &  (cm)      &        &       \\
		\hline
		Abastumani       & Georgia     & 70    & {\LARGE $\diamond$} & dark green \\
		Aoyama-Gakuin    & Japan       & 35    & {\LARGE $\diamond$}  & cyan \\
		ARIES            & India       & 130   & {\large $\times$}    & blue\\
		Crimean          & Russia      & 70    & $\square$   & magenta\\
		Lulin            & Taiwan      & 40    & {\LARGE $\circ$}     & cyan\\
		Mt.~Maidanak     & Uzbekistan  & 60    & {\LARGE $\circ$}     & blue\\
		Osaka Kyoiku     & Japan       & 51    & {$\triangle$} & magenta\\
		Roque de los     & Spain & 260   & {\LARGE $\diamond$}  & blue\\
                Muchachos$^a$    &               &                      &      \\
		Rozhen           & Bulgaria    & 50/70 & {\LARGE $\diamond$}  & black\\
		Rozhen           & Bulgaria    & 200   & $\square$   & black\\
		San Pedro Martir & Mexico      & 84    & {\large $+$}         & green\\
		Siena            & Italy       & 30    & {\large $+$}         & black\\
		Sirio            & Italy       & 25    & {\large $\times$}    & orange\\
		St.~Petersburg   & Russia      & 40    & $\square$   & green\\
		Teide$^b$        & Spain       & 120   & {$\triangle$} & blue\\
		Tijarafe         & Spain       & 40    & {\large $+$}         & red \\
		\hline
	\end{tabular}
$^a$ Nordic Optical Telescope (NOT)\\
$^b$ STELLA-I telescope
\end{table}

\begin{figure}
	\includegraphics[width=\columnwidth]{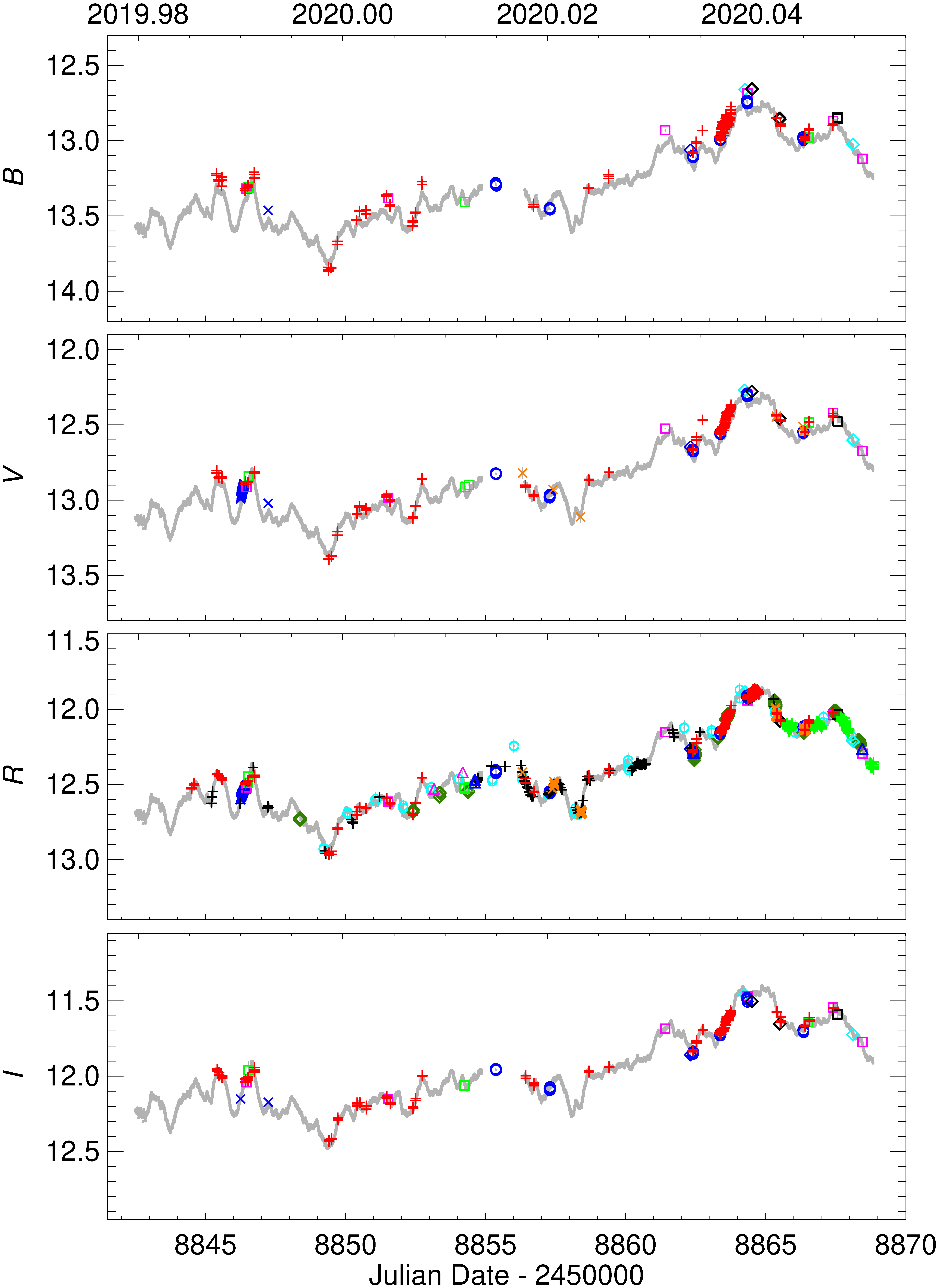}
    \caption{Light curves of S5~0716+714 in $BVRI$ built with data from the WEBT Collaboration. Different symbols and colours distinguish different datasets as explained in Table~\ref{list}. In all panels, the {\it TESS} light curve is plotted in grey, highlighting the excellent agreement between space- and ground-based data. }
    \label{webt}
\end{figure}

Table~\ref{list} shows the WEBT observatories which supported the {\it TESS} monitoring of S5~0716+714 with multiband observations. 
These were mainly done in the Johnson-Cousins $BVRI$ bands, with a few data acquired in Johnson's $U$ and Sloan's $z$ bands.
Data reduction was performed with standard methods. Calibration of the source was obtained using the photometric sequences by \citet{villata1998a} in $BVR$, \citet{ghisellini1997} in $I$, \citet{gonzalez2001} in $U$; $z$ data were calibrated with Pan-STARRS PS1 data\footnote{https://panstarrs.stsci.edu/}.
In total, the WEBT collected 2200 data points, 1491 of which in the $R$ band.

We assembled the different datasets and carefully inspected the WEBT multiband light curves night per night. A handful of outliers were removed. We had to apply small shifts to a couple of datasets in $B$ band and to one dataset in $R$ band  to align them to the common trend. In this way we obtained homogeneous light curves. These are plotted in Fig.~\ref{webt} and reveal that the source was in a fairly bright state when compared to past observations \citep[e.g.][]{villata2000,raiteri2003,hagen2006,villata2008}.
They also show that the variability amplitude increases from red to blue, as usually found in BL Lac objects.
The differences between the maximum and minimum magnitude are 0.974, 1.111, 1.125, 1.212 in $IRVB$, respectively.

Fig.~\ref{webt} also shows the {\it TESS} fluxes transformed into ``equivalent" magnitudes according to the formula: 
$\rm mag_{\it TESS}=-2.5 \, \log $ (SAP\textunderscore FLUX)$\rm + mag_0$, where the zero-point mags are: $\rm mag_0 = 21.30, \, 20.85, \, 20.42, \, 19.96$ in $BVRI$, respectively. These values have been set empirically to make the {\it TESS} light curve match the overall behaviour of the WEBT ones, and represent average offsets over the whole light curves.
The comparison between the space- and ground-based data reveals an excellent agreement. The differences in variability amplitude in $V$ and especially in $B$ can be ascribed to the fact that the {\it TESS} spectral response function\footnote{https://heasarc.gsfc.nasa.gov/docs/tess/the-tess-space-telescope.html} is very wide and covers the whole spectral range of the $z$ and $I$ bands, most of the spectral range of the $R$ band, but overlaps only in part with the $V$-band transmission curve, and has practically no overlap with that of the $B$ band.

\begin{figure}
	\includegraphics[width=\columnwidth]{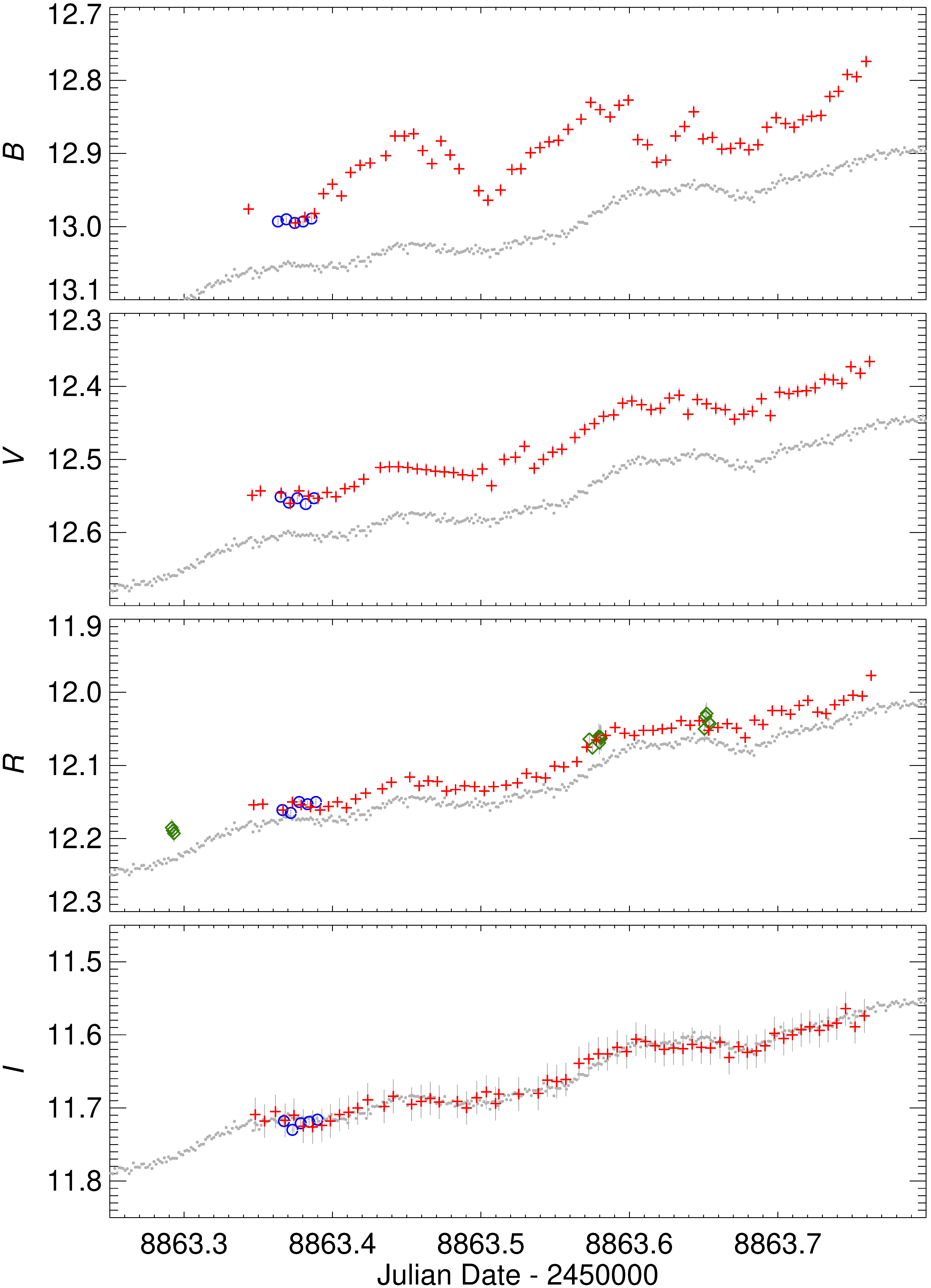}
    \caption{A zoom into Fig.\ \ref{webt} to highlight the source brightness and spectral variability on $\rm JD-2450000=8863$. Densely-sampled observations were done at the Tijarafe Observatory (red plus signs) with contributions by the Mt.\ Maidanak  (blue circles) and Abastumani (green diamonds) observatories. }
    \label{webt_idv}
\end{figure}

A better view on the brightness and spectral short-term variability is given in Fig.\ \ref{webt_idv}, which presents a zoom into the multiband light curves on $\rm JD-2450000=8863$. 
The {\it TESS} light curve is reported here with the same scaling as in Fig.\ \ref{webt}. As mentioned above, the zero points were set by considering the bulk behaviour over the whole {\it TESS} observing period. A comparison with Fig.\ \ref{webt} shows that on $\rm JD-2450000=8863$ the source was in a relatively high state, so we expect differences between ground and space data, more pronounced toward the blue, i.e.\ moving away from the {\it TESS} passband.
Indeed, the densely-sampled intra-night observations at the Tijarafe Observatory show an excellent agreement with the {\it TESS} light curve in $I$ band, while ground-based data in $R$, $V$ and $B$ bands show increasing deviations from the space data and increasing oscillation amplitudes. The offset between space and ground data is of the order of $\sim 0.02$ mag in $R$ band, and ranges between 0.04 and 0.08 mag in $V$ band and between 0.04 and 0.15 mag in $B$ band.
We also notice that the brightness rising phase starting at $\rm JD-2450000 \sim 8863.5$ in $B$ band possibly leads the corresponding variation in $V$ band; lags in $R$ and $I$ bands are difficult to assess because of the smaller variability.

\begin{figure}
	\includegraphics[width=\columnwidth]{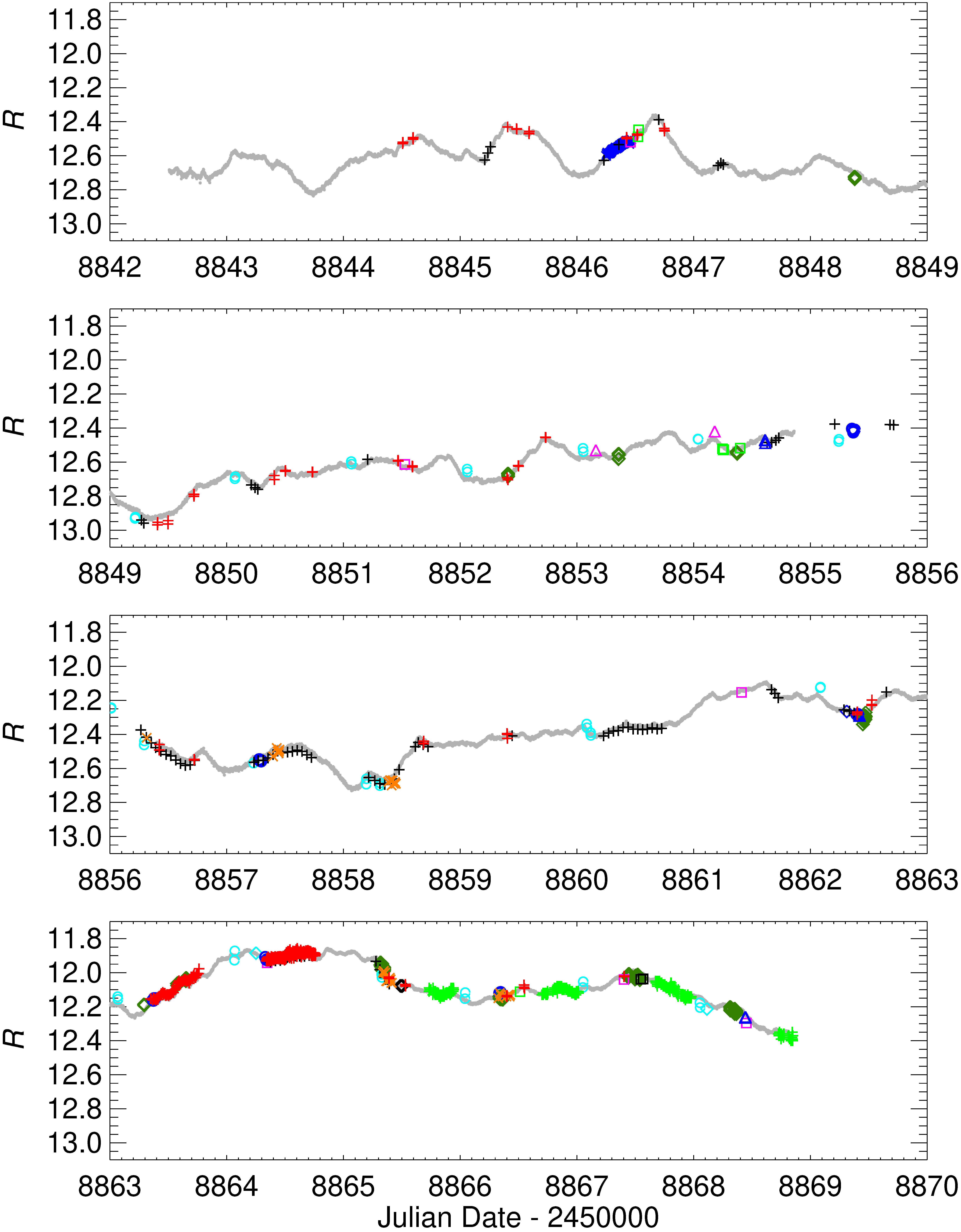}
    \caption{The WEBT $R$-band and {\it TESS} light curves of S5~0716+714 shown in Fig.~\ref{webt} folded in time intervals of one week.}
    \label{folded}
\end{figure}

Microvariability on the whole {\it TESS} observing period can be appreciated in Fig.~\ref{folded}, where we show again the WEBT $R$-band light curve folded in time intervals of one week, superposed to the {\it TESS} light curve.
The WEBT sampling is sparse in the first week and becomes denser thereafter, especially in the last week. 
Long intra-night sequences were made at the San Pedro Martir, Siena, and Tijarafe observatories. These sequences follow very closely the trend traced by the 2-min cadence {\it TESS} data.
The good agreement between the ground and space data also provides a validation of the {\it TESS} photometry.

\begin{figure}
	\includegraphics[width=\columnwidth]{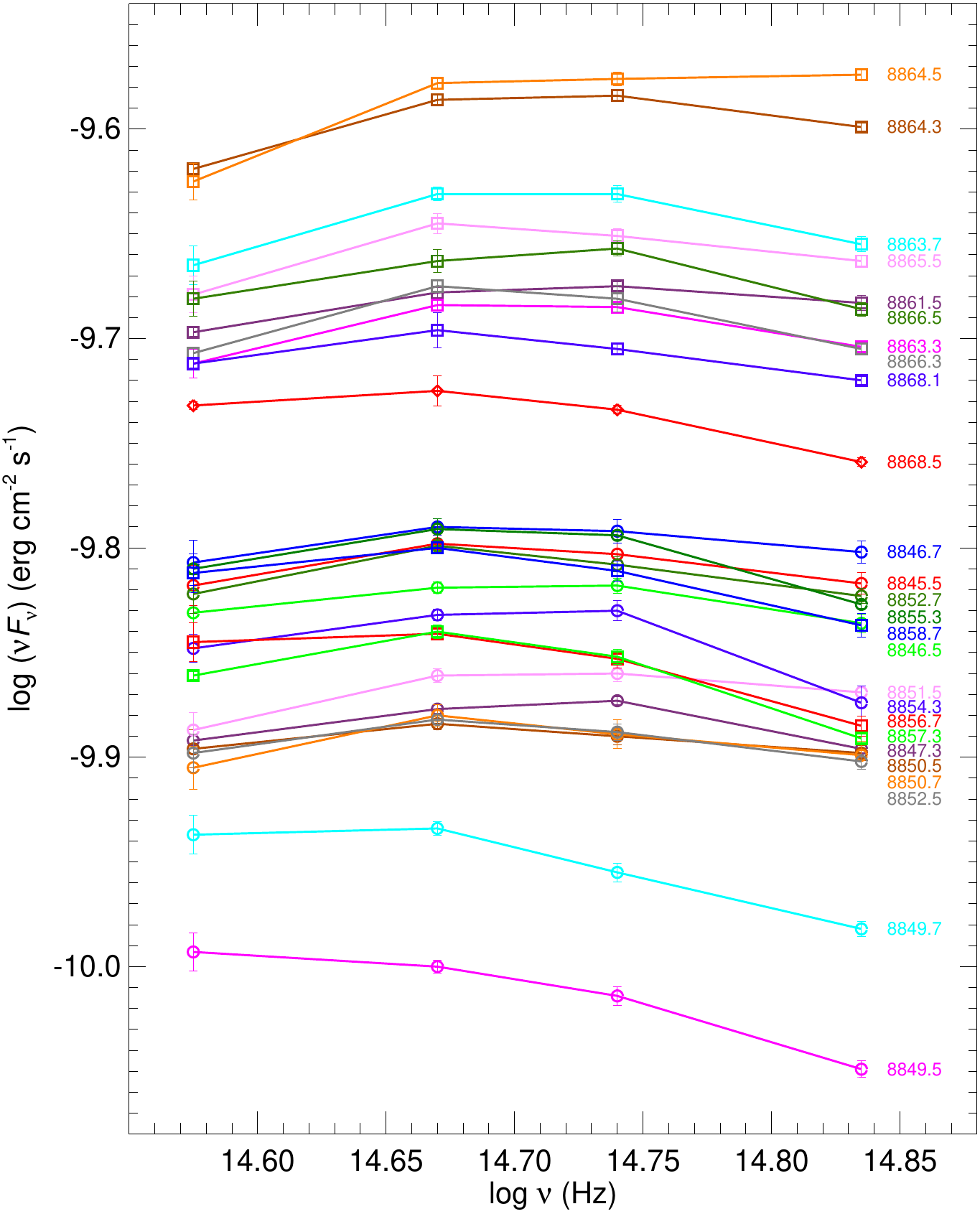}
    \caption{Optical SEDs of S5~0716+714 at various epochs (indicated on the right as $\rm JD-2450000$) during the period of  {\it TESS} observations. Each SED has been built with multiband data acquired by the WEBT Collaboration within 0.2 d.}
    \label{spettri}
\end{figure}

\begin{figure*}
	\includegraphics[width=12cm]{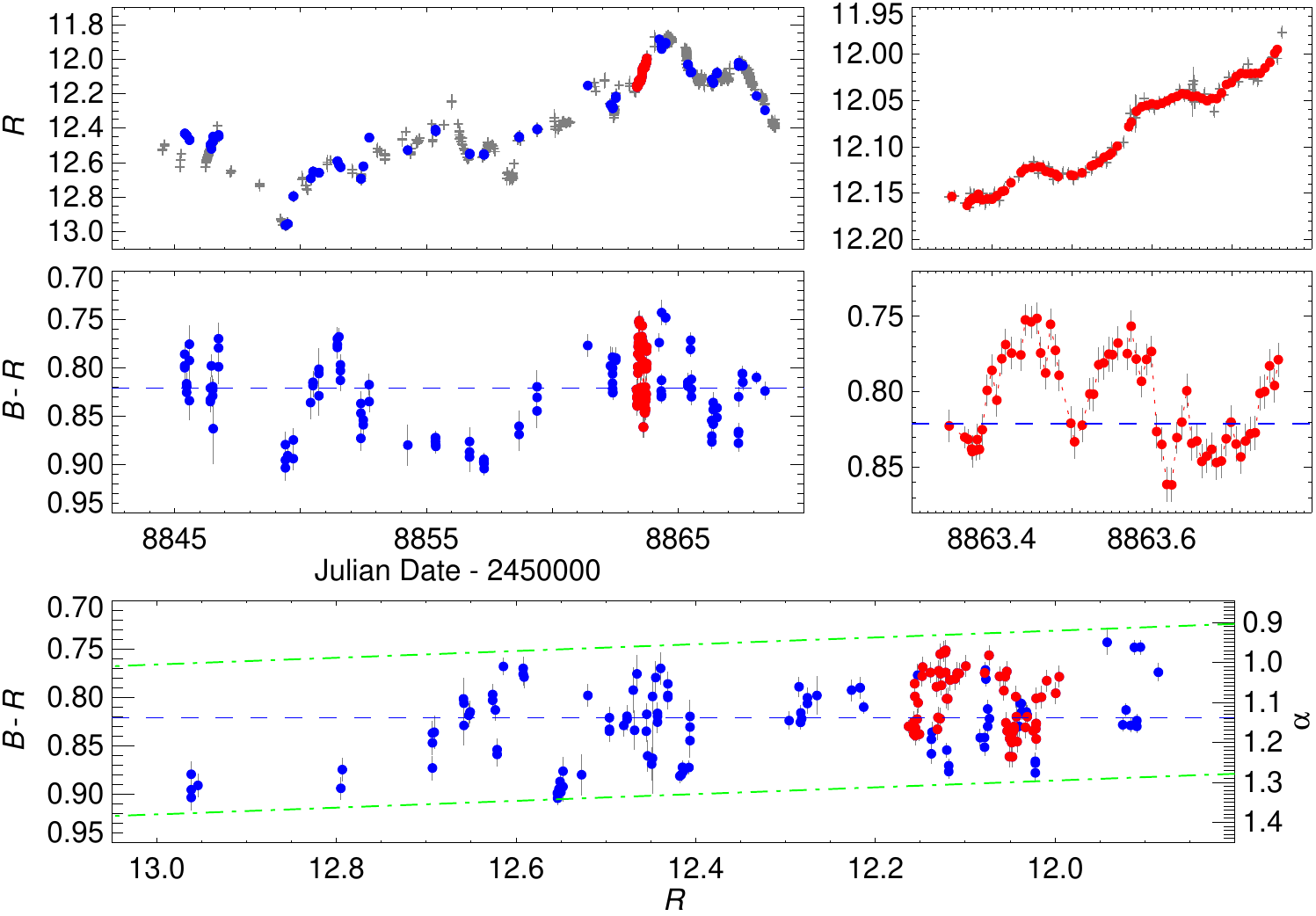}
    \caption{Top: the WEBT $R$-band light curve (grey plus signs) over the whole {\it TESS} observing period (left) and on $\rm JD-2450000=8863$ (right). The blue and red dots mark the data points used to build $B-R$ colour indices.
Middle: the $B-R$ colour indices obtained by coupling $B$ and $R$ WEBT data taken by the same telescope within 30 minutes.
Bottom: the same indices versus brightness. In all plots, red dots refer to data points corresponding to $\rm JD-2450000=8863$. The green dot-dashed lines follow the $B-R$ maxima and minima trends and define a mild bluer-when-brighter behaviour ascribable to the long-term flux variability. 
The right vertical axis gives the spectral index $\alpha$ ($F_\nu \propto \nu^{-\alpha}$) calculated between the $R$ and $B$ de-reddened flux densities. Horizontal dashed blue lines indicate the average $B-R$ (and $\alpha$) value.}
    \label{colori}
\end{figure*}

S5~0716+714 is a distant BL Lac object, therefore its optical emission is completely dominated by the non-thermal synchrotron emission from the jet. 
Fig.~\ref{spettri} shows the optical spectral energy distributions (SEDs) in the usual $\log (\nu F_\nu)$ versus $\log \nu$ representation at various epochs during the {\it TESS} pointing period. 
SEDs have been built with data from the WEBT Collaboration binned in time intervals of 0.2 d, i.e.\ the data points belonging to a given SED represent averages of data acquired within 4.8 h. 
Magnitudes have been corrected for Galactic extinction using the values of the NASA/IPAC Extragalactic Database\footnote{https://ned.ipac.caltech.edu/} (NED), and then transformed into flux densities adopting the zero-point values of \citet{bessell1998}.

The SEDs show a variety of spectral shapes and there is not a clear trend with brightness. However, the faintest states display a soft spectrum, suggesting that the synchrotron frequency peak is in the red part of the spectrum, while the brightest states indicate that the peak has moved toward the blue part. 

To further analyse the optical spectral variability, we plot in Fig.~\ref{colori} the $B-R$ colour indices obtained by coupling $B$ and $R$ data points taken by the same telescope within 30 min. This represents a stricter constraint than that used to build the SEDs. We obtained 180 colours, with an average value 0.82 and standard deviation 0.04. The figure shows that spectral changes on short time-scales, even within the same night, span almost the whole range of $B-R$ values.
This means that the long-term flux oscillations are much less chromatic than the fast flux changes. 
The strong spectral variations on short time-scales are not simply correlated with the flux, as the data acquired on $\rm JD-2450000=8863$ demonstrate.
There, a bluer-when-brighter trend can be seen if we correlate the colour index with the small and fast flux changes that are superposed on the quasi-linear increasing base level. We will come back to this point in Section~\ref{seccro}.

The bulk behaviour of $B-R$ versus $R$ shows only a mild bluer-when-brighter trend, ascribable to the long-term flux variability, with $\Delta (B-R) / \Delta R \sim 0.035$.
This has been inferred by defining a stripe whose boundaries follow the $B-R$ maxima and minima trends.
In comparison, the ratio between the maximum colour difference and the $R$-band range spanned by the points used to build colours is 0.15, i.e.\ more than 4 times larger.

From the $B$ and $R$ de-reddened flux densities we derive the optical spectral index $\alpha$ ($F_\nu \propto \nu^{-\alpha}$). It ranges from 0.95 to 1.34, with a mean value of 1.14 and standard deviation 0.09.
All $\alpha$ values below unity, which indicate a hard spectrum, belong to bright states, while the faintest states are characterized by the largest $\alpha$ values, i.e.\ the softest spectra. This confirms what was already noticed above for the SEDs.
Moreover, from both spectra and colour indices it can be seen that spectral hardness is not strictly depending on brightness (apart from the most extreme states), but it most likely depends on the very fast changes, according to the harder-when-brighter relation noted above and already reported by \citet{ghisellini1997} and \citet{raiteri2003} for the same source in earlier periods.
This double (long and short-term) chromatism is rather common in BL Lac objects, as shown e.g.\ for BL Lacertae during WEBT campaigns \citep{villata2002,villata2004,papadakis2007}.

\section{Observations by {\it Swift}}
\label{secx}
The multiband WEBT data presented in Section~\ref{secwebt} provide spectral information in the optical wavelength range. To extend the spectral information to higher frequencies, we asked for ToO observations by the {\it Neil Gehrels Swift Observatory} during the {\it TESS} pointing. 
{\it Swift} can provide data in the optical ($v$, $b$, and $u$ bands) and ultraviolet ($w1$, $m2$, and $w2$ bands) with the 30-cm Ultraviolet/Optical telescope \citep[UVOT,][]{roming2005}, and at X-rays (0.3--10 keV) with the X-ray Telescope \citep[XRT,][]{burrows2005}.
We processed the satellite data with the {\small HEASoft} software and calibration data available at NASA's High Energy Astrophysics Science Archive Research Center\footnote{https://heasarc.gsfc.nasa.gov/} (HEASARC), following standard procedures.

\begin{figure}
	\includegraphics[width=\columnwidth]{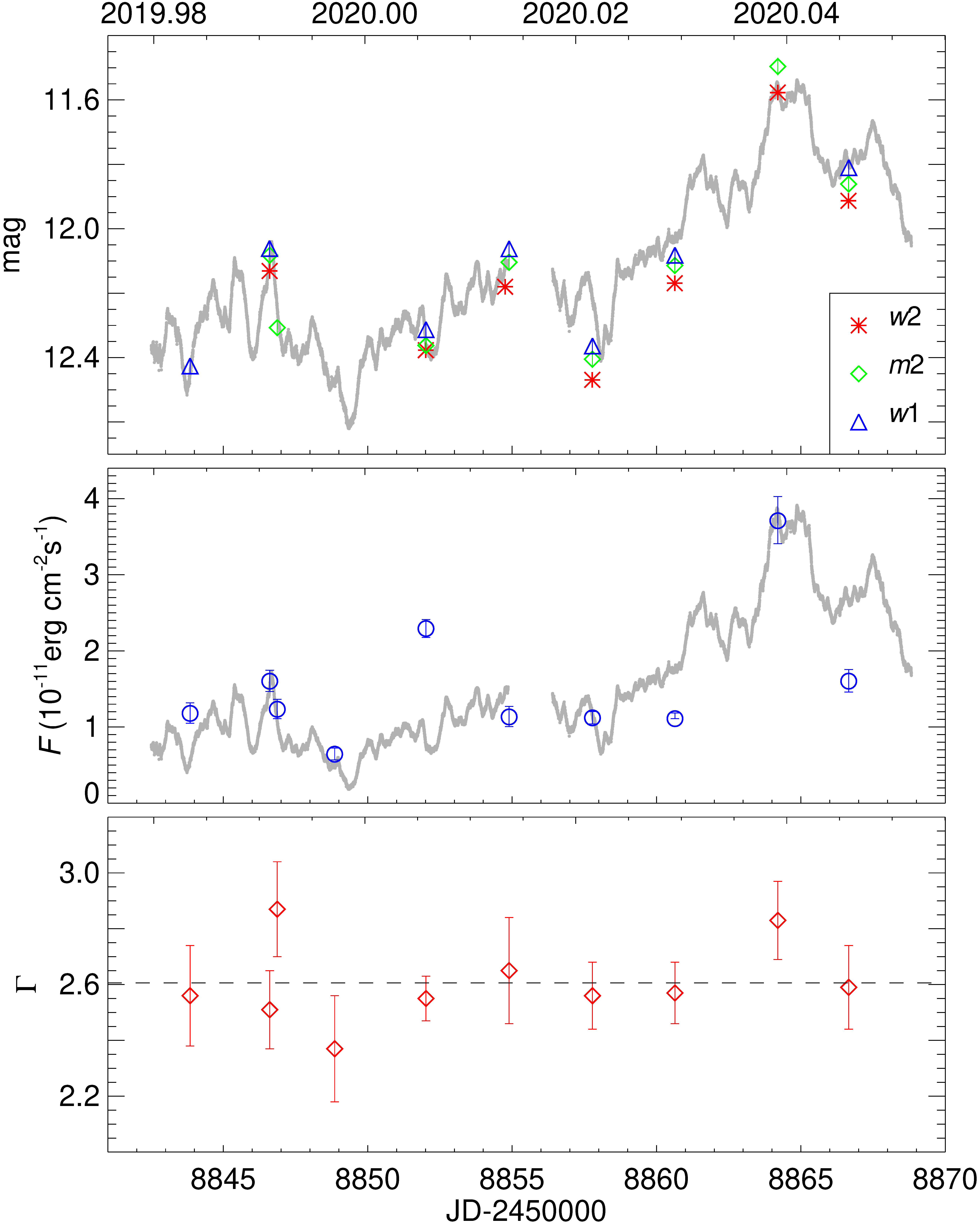}
    \caption{Top: Ultraviolet magnitudes of S5~0716+714 obtained from the {\it Swift}-UVOT observations during the {\it TESS} pointing. Different colours and symbols identify the various filters, as shown in the inset. The grey line represents the {\it TESS} fluxes $F$ scaled as $-2.5 \, \log F + 20.1$.
Middle: The {\it Swift}-XRT de-absorbed fluxes in the 0.3--10 keV spectral range obtained by fitting the X-ray spectra with an absorbed power-law model with the hydrogen column density fixed to the Galactic value $N_{\rm H}=2.88 \times 10^{20} \rm \, cm^{-2}$. The grey line shows the {\it TESS} fluxes scaled as $F/450-2$.
Bottom: The photon index $\Gamma$ corresponding to the same model fits.}
    \label{swift}
\end{figure}

The S5~0716+714 counts were extracted from the UVOT images within a circle of 5 arcsec radius centred on the source, and background counts within a 20 arcsec radius circular region in the surroundings. We checked that the source is not affected by small scale sensitivity problems. The UVOT ultraviolet light curves are plotted in Fig.~\ref{swift}. 

\begin{figure}
	\includegraphics[angle=-90,width=\columnwidth]{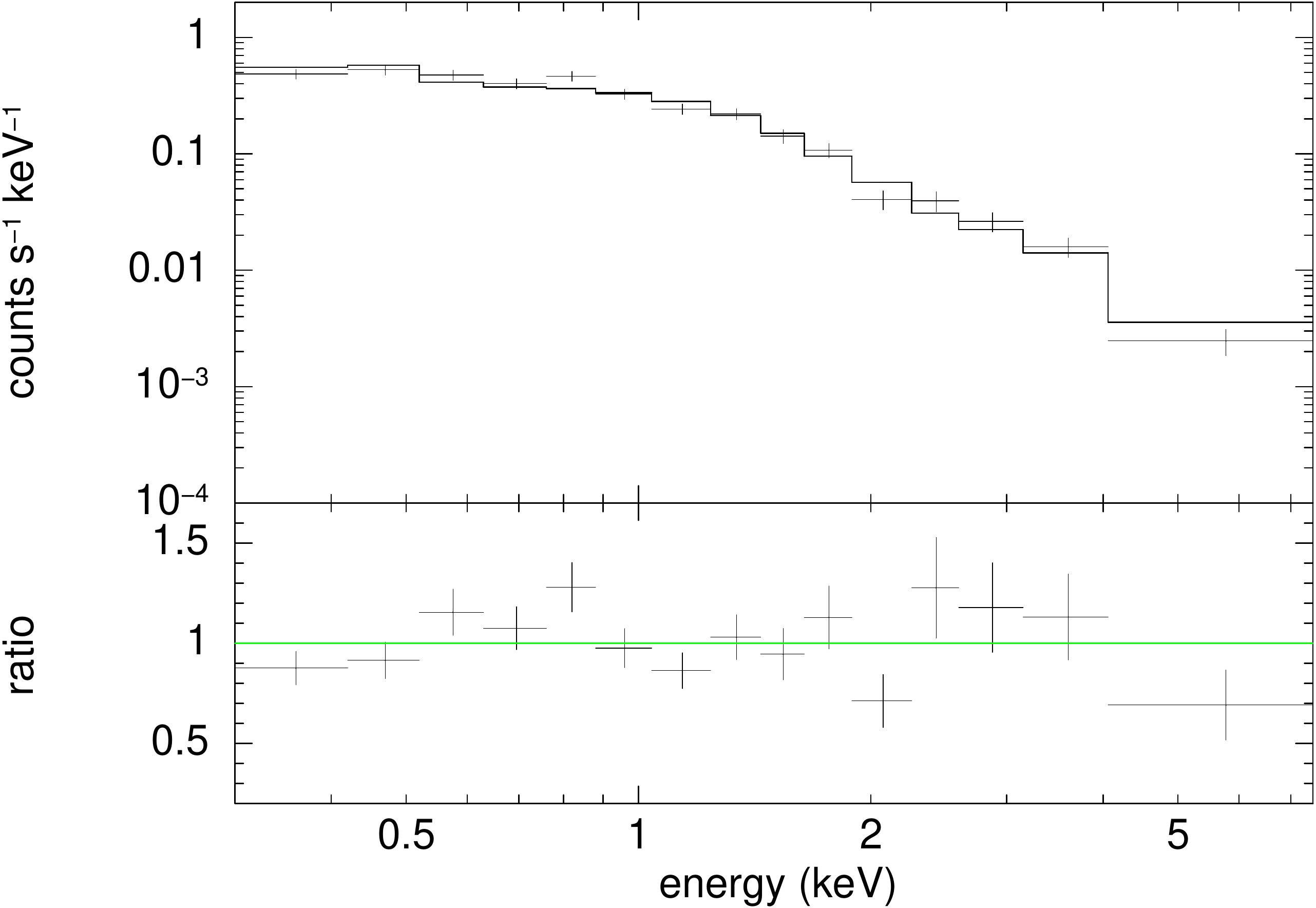}
    \caption{Top: The binned X-ray spectrum of S5~0716+714 acquired by {\it Swift}-XRT on 2020 January 3. It is fitted with an absorbed power-law model, with the hydrogen column density fixed to the Galactic value $N_{\rm H}=2.88 \times 10^{20} \rm \, cm^{-2}$.
Bottom: The ratio between data and model.}
    \label{xrt}
\end{figure}

The same figure also shows the source fluxes in the 0.3--10 keV energy range. They were obtained by fitting the XRT data with an absorbed power-law model using the {\small XSPEC} software.
The hydrogen column density was fixed to the Galactic value $N_{\rm H}=2.88 \times 10^{20} \rm \, cm^{-2}$ derived\footnote{https://www.astro.uni-bonn.de/hisurvey/AllSky\textunderscore profiles/} from the Effelsberg-Bonn \ion{H}{I} Survey \citep{winkel2016}. Because of the low number of counts we used the Cash statistics \citep{cash1979}.
As can be seen in Fig.~\ref{xrt}, which displays the XRT spectrum of 2020 January 3 ($\rm JD=2458852.02$) together with the best-fit absorbed power-law, the model gives a good description of the data.

The comparison between the source behaviour in the ultraviolet and X-ray bands and the {\it TESS} light curve reveals some interesting features. The ultraviolet magnitudes fairly agree with the {\it TESS} fluxes properly scaled as $-2.5 \, \log F + 20.1$. In line with what has been already noticed for the optical bands in Section~\ref{secwebt}, the variability amplitude in ultraviolet appears larger than in the {\it TESS} band. 
The X-ray light curve sometimes correlates with the optical one, while sometimes it does not. In particular, the X-ray flux is higher than expected on $\rm JD-2450000=8852$ and lower on 8860 and 8866. The former case suggests that the X-ray emission can receive an additional contribution from a source that does not affect the optical output. The latter cases may indicate that the mechanism which leads to the long-term oscillation of the baseline optical flux level is much less effective at X-rays.

The photon index $\Gamma$, shown in Fig.~\ref{swift}, is affected by large uncertainties and is in general compatible with its average value $\Gamma=2.61 \pm 0.15$. It never goes below $\Gamma=2$, which means that the X-ray spectrum of S5~0716+714 is always soft. This in turn implies that the 0.3--10 keV flux receives a dominant emission contribution from the synchrotron process rather than from the inverse-Compton process that is responsible for the emission at hard X-ray and $\gamma$-ray energies. 
We note that the dimmest X-ray level corresponds to the hardest X-ray spectrum and the brightest X-ray level is characterized by a softer spectrum than average.
Therefore, it seems that when the synchrotron emission is fainter, the larger weight of the inverse-Compton contribution makes the spectrum harder.

We finally note the fast brightness decrease in the $m2$ band on 2019 December 29 ($\rm JD-2450000=8846.6$--8846.9), with a change of more than 0.2 mag (from $m2 = 12.091 \pm 0.025$ to $12.314 \pm 0.024$) in 0.26 d.
This decrease is consistent with the behaviour in the optical band, as highlighted by the {\it TESS} light curve.
Contemporaneously, the X-ray flux decreased by about 30 per cent. We also checked for significant flux changes within the same ObsID. On 2020 January 6 ($\rm JD-2450000=8854.6$--8854.9) we detected an increase of the source X-ray count rate by a factor $\ga 2$, from $(0.203 \pm 0.048) \rm \, cts \, s^{-1}$ to $(4.365 \pm 0.034) \rm \, cts \, s^{-1}$ in 0.27 d with a $\sim 4 \sigma$ significance. In the same epoch, we have a double measurement in the UVOT $w2$ band, showing a brightening from $(12.271 \pm 0.027) \rm \, mag$ to $(12.145 \pm 0.023) \rm \, mag$ in nearly the same time span. 

\section{Time series analysis of {\it TESS} data}
\label{sectsa}

\subsection{Power spectrum}
\label{secpsd}
We now investigate the processes at work in the {\it TESS} light curve by means of the periodogram. The periodogram is obtained by performing the Fourier transform  of a time series and gives an estimate of its power spectral density (PSD, also named power spectrum) \citep[see][for a review]{vanderplas2018}. 
When modelled as a power law: $P(f) \propto f^{-\alpha}$, where $f$ is the frequency, the spectral index depends on the noise process:
$\alpha=0$ characterizes white noise,
$\alpha=1$ flicker noise, and 
$\alpha=2$ red (or random-walk) noise. 
Because of the presence of a gap in the data, we estimate the PSD with methods for unevenly sampled datasets, based on discrete Fourier transform \citep{lomb1976,scargle1982,horne1986,press1992}.

\begin{figure}
	\includegraphics[width=\columnwidth]{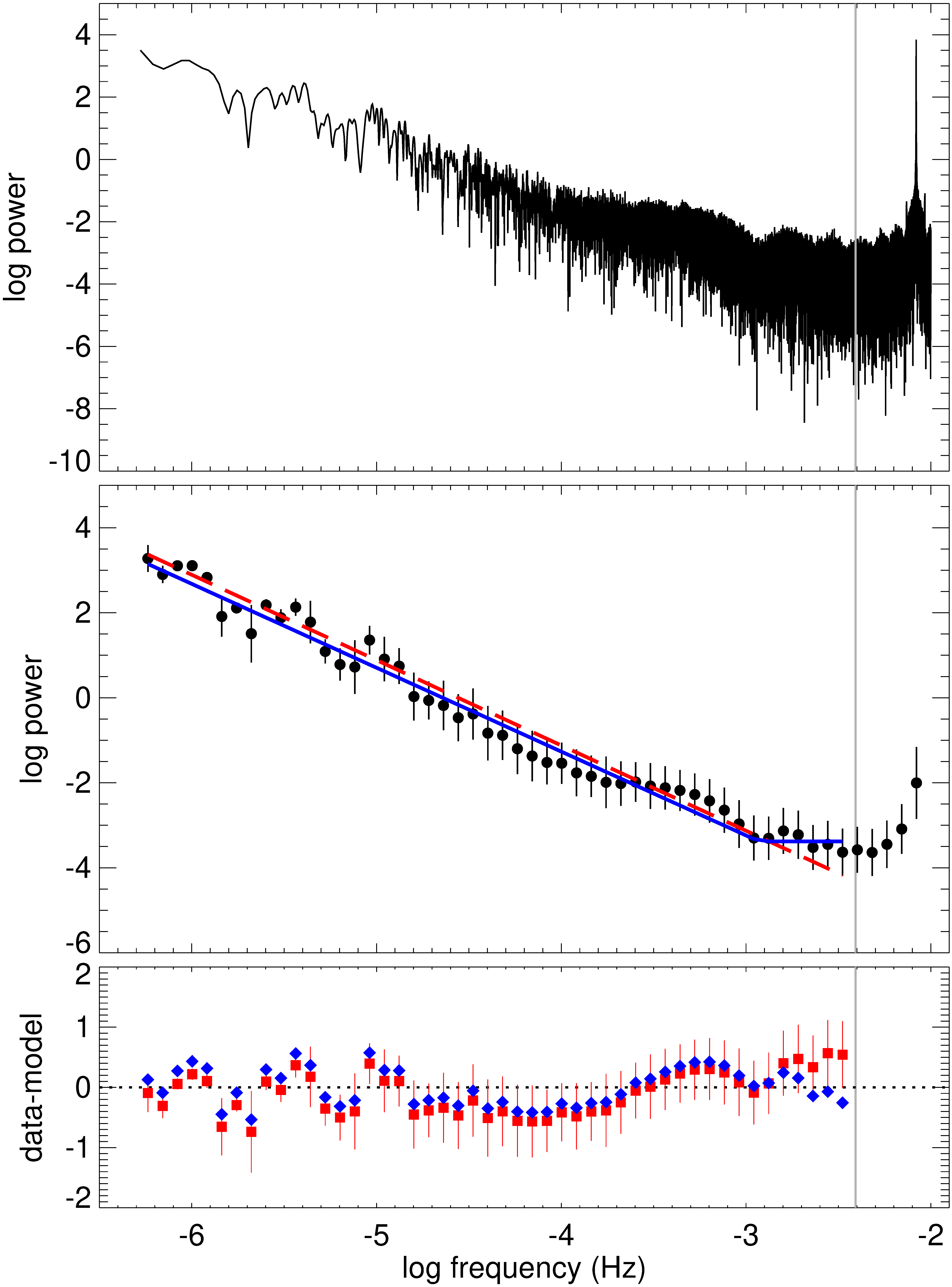}
    \caption{Top: The periodogram built with the {\it TESS} fluxes of S5~0716+714 versus frequency in the log-log representation. An oversampling factor of 5 has been used.
    Middle: Binned periodogram; errors are calculated as standard deviations of the log power in the bins; the red dashed line represents the best-fit power-law model obtained on data up to the Nyquist critical frequency marked with a grey vertical line; the blue solid line highlights the broken power-law with high-frequency slope set to zero.
    Bottom: Residuals (data minus model) for both fits: red squares refer to the simple power-law case, blue diamonds to the broken power-law case.}
    \label{psd}
\end{figure}

Fig.~\ref{psd} shows the periodogram versus frequency in a log-log plot. An oversampling factor of 5 was used to increase the number of frequencies for which the power is calculated \citep{press1992}.
The periodogram displays a roughly linear decreasing trend for frequencies lower than the Nyquist critical frequency
$f_c=N/(2 \, \Delta T) \approx 0.004 \rm \, Hz$ ($\log f_c \approx -2.41$), where $N$ is the number of points of the light curve and $\Delta T$ is the length of the observing period.
This frequency corresponds to a time interval of about 4 min, i.e.\ twice the sampling one. 
For higher frequencies the periodogram inverts the trend and turns up. This is the effect of the signal due to the 2-min time separation between data points.

Because of the sampling in frequency, the higher-frequency range is much more represented. Therefore, to estimate the spectral index of the power-law model, we bin the periodogram in steps of $\Delta \log f=0.08$ up to $f_c$ and then perform a least-squares linear fit.
The slope/spectral index is $\alpha=2.01 \pm 0.04$, which confirms the red noise nature of the process already noticed in Section~\ref{sectess}. The reduced chi-squared goodness of fit is $\chi^2/{\rm(dof)}=0.955$, for dof=46 degrees of freedom, and indicates a good fit. A Kolmogorov--Smirnov (KS) test gives a probability of 99.4 per cent that the points of the binned log-log periodogram and those of the corresponding power-law model fit are drawn from the same distribution.

The lower panel of Fig.~\ref{psd} shows the residuals (data minus model). They allow us to better recognize the deviations of the data from the model. While at 
$\log f \la -4.9$ there is some dispersion around 0, at higher frequencies the deviations from the model appear to follow a path. 
Notwithstanding the large errors, this suggests that a power-law model for the PSD may be an oversimplification. In particular, the last residual points indicate underfitting, in agreement with the flattening of the periodogram above $\log f \sim -3$.
This may be due to red noise turning into instrumental white noise above some break frequency, as found in the {\it Kepler} light curves of some AGN \citep{wehrle2013,sasada2017,wehrle2019}.
Therefore, we perform a second fit using a broken power-law, forcing the slope of the higher-frequency linear fit to be zero. The result is a statistically worse fit, with $\chi^2/{\rm(dof)}=1.48$. 
The best-fit break frequency is $\log f=-2.93$ (about 14 min), and the slope/spectral index of the lower-frequency linear fit $\alpha=1.97 \pm 0.16$, not very far from the simple power-law case (see Fig.~\ref{psd}). The KS test leads to the same probability as before (within the sixth decimal digit).

In summary, the evaluation of the PSD of S5~0716+714 obtained from the {\it TESS} 2-min cadence light curve indicates that the main variability process operating is essentially red noise,  which possibly becomes instrumental white noise on time-scales shorter than about 14 min. However, the fine structure of the periodogram, which presents some bending, reveals a more complex situation, whose explanation is hard to figure out. Complex shapes of the blazar optical PSD were also found by other authors \citep{edelson2013, sasada2017,goyal2018}.

\begin{figure}
	\includegraphics[width=\columnwidth]{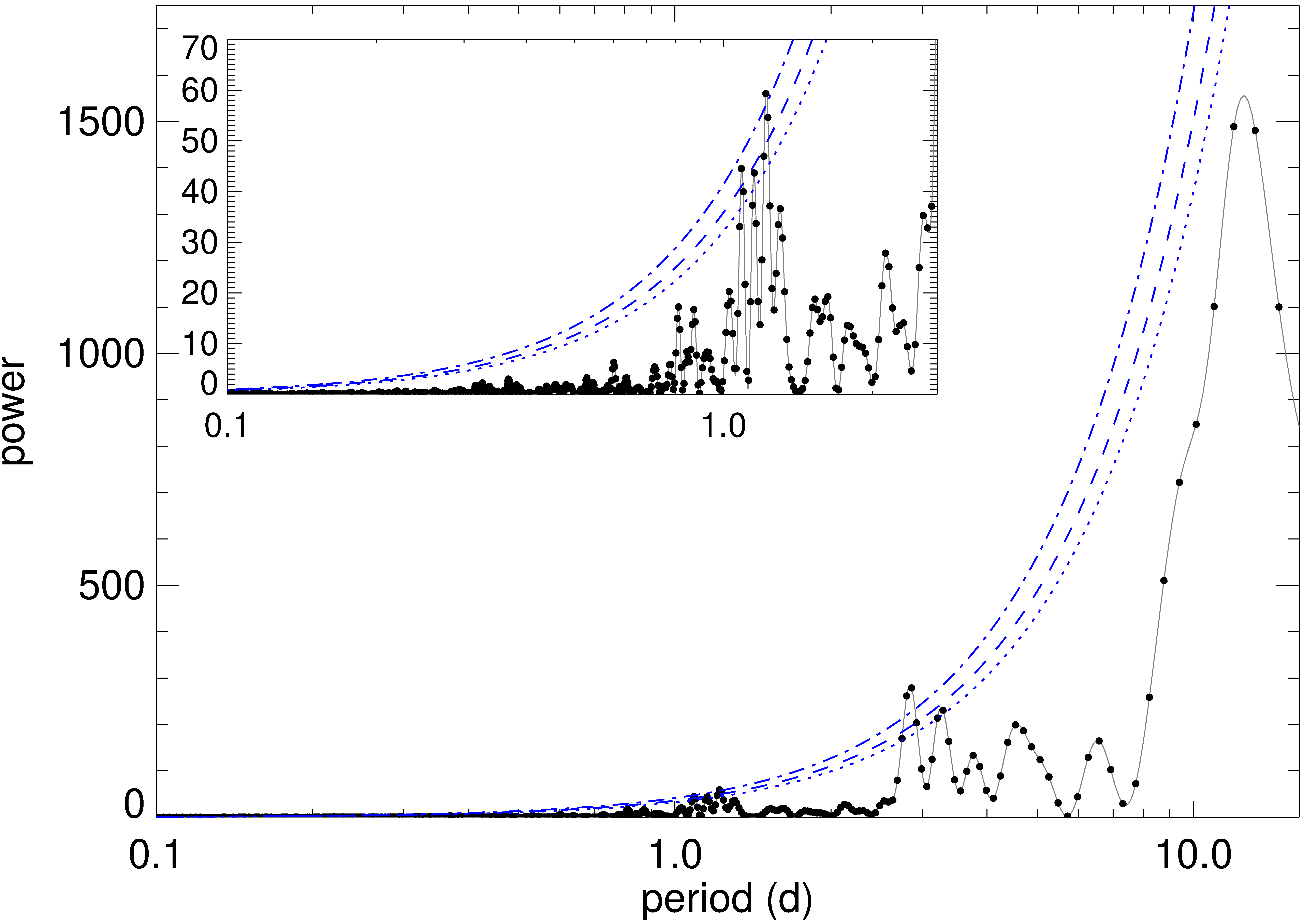}
    \caption{The periodogram of S5~0716+714 built with the {\it TESS} data versus period. Black points refer to the periodogram shown in Fig.~\ref{psd}, which is sampled in frequency and thus has few points at long periods. The grey line is the result of calculating the periodogram by sampling in period instead of frequency. It is shown to guide the eye through the low-frequency peaks. 
The inset displays a zoom on the 0.1--2.7 d period range. 
In both plots the blue lines indicate the significance levels of 95 (dotted), 99 (dashed) and 99.9 (dot-dashed) per cent. }
    \label{powert} 
\end{figure}

Besides revealing the type of process determining the source variability, the periodogram can be used to investigate the presence of sinusoidal components in a light curve, i.e.\ of periodicities. These will produce peaks in the PSD at the corresponding frequencies/periods.
Indeed, some of the PSD peaks in Fig.~\ref{psd} may in principle be due to periodic signals.
For convenience, in Fig.~\ref{powert} we show the periodogram versus time. A series of peaks are present, whose power increases with period, as expected for red noise and already visible in Fig.~\ref{psd}. 
The evaluation of peak significance in presence of red noise is a crucial issue that has been tackled in different ways \citep{benlloch2001,vaughan2005,vaughan2010,vanderplas2018}.
We adopted the method presented by \citet{vaughan2005}. 
We first fit the unbinned log-log periodogram of Fig.~\ref{psd} with a power-law model up to the Nyquist frequency. 
Then we add the logarithm of 
$\gamma_\epsilon/2=-\ln[1-(1-\epsilon)^{1/M}]$ to the model, 
where $M$ is the number of points of the periodogram that have been used for the power-law fit, $\epsilon$ is the false alarm probability, so that $1-\epsilon$ is the significance level for a detection, and $\gamma_\epsilon$ is the corresponding confidence limit.
In Fig.~\ref{powert} we plot significance levels of 95, 99, and 99.9 per cent.
Among the several peaks displayed in Fig.~\ref{powert}, some of which appear very close in period, only a few seem to be marginally significant. They correspond to periods of 1.1--1.2 d and 2.8--3.3 d. This suggests that in the time span that we are considering, there are events which may repeat regularly. No significant peak is found below 1 d, possibly meaning that the very fast flux changes that we noticed in the light curves are not part of a regular process.
However, the fact that some of the periodogram signals present very similar periods also suggests that we may see the signature of quasi-periodicities, QPOs. In this case, the power of the QPO is spread over multiple frequencies and this reduces its significance.
In the following section, we will use another method to check whether the above QPOs can be confirmed and to better characterize the short-term variability. 

\subsection{Auto-correlation function}
\label{secacf}
One popular time series analysis technique is the discrete correlation function \citep{edelson1988,hufnagel1992}, which quantifies the correlation between two unevenly-sampled datasets. 
When these coincide, we obtain an auto-correlation function (ACF).
Deep minima in the ACF highlight characteristic variability time-scales, while periodicities produce strong ACF peaks, i.e.\ with values close to 1, for time separations $\tau$ corresponding to the period and its multiples.
While the PSD is sensitive to sinusoidal signals in the data train, the ACF is strongly affected by peaks and dips of the light curve.

\begin{figure*}
	\includegraphics[width=12cm]{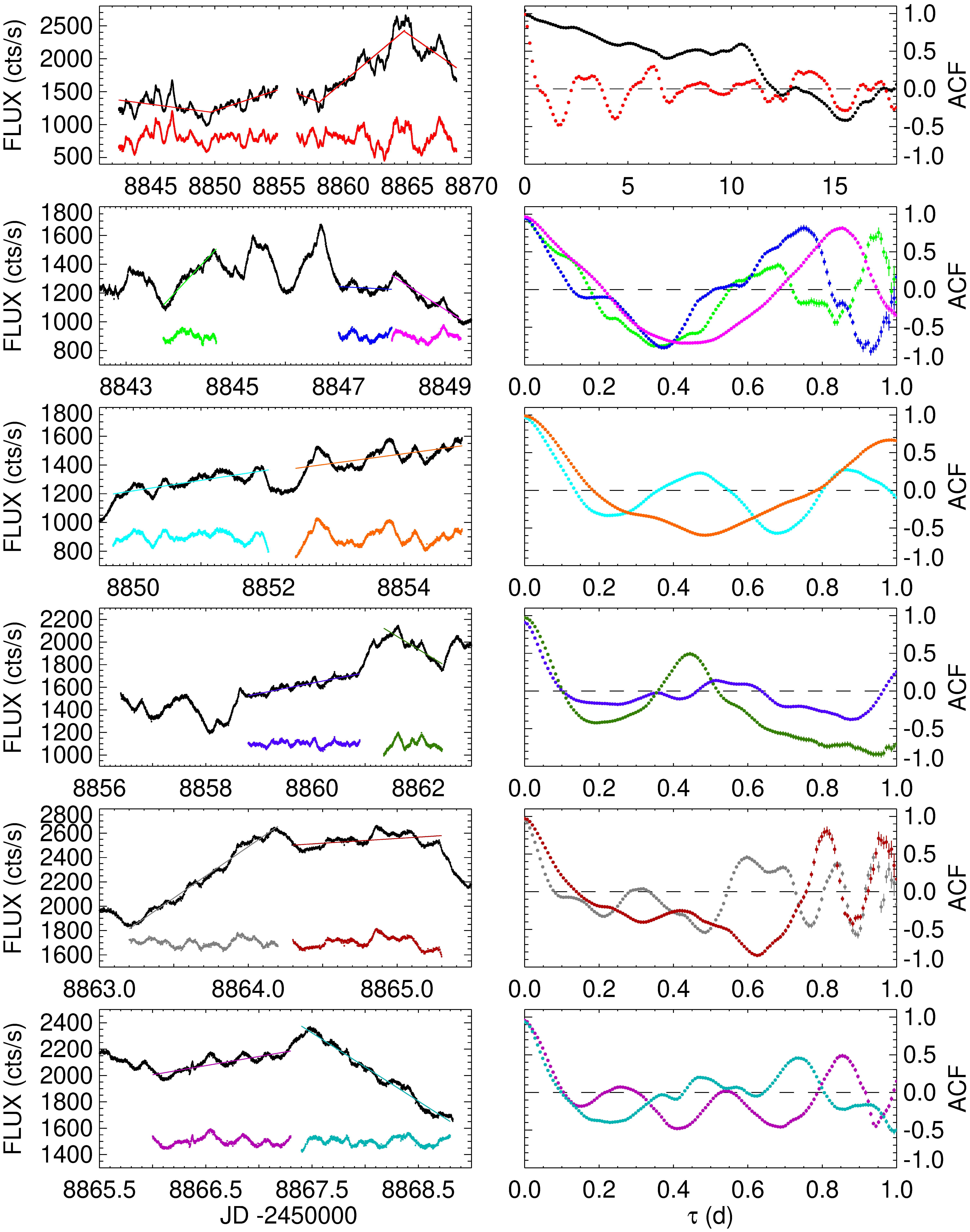}
    \caption{Left panels: {\it TESS} observed light curve (black) and linearly de-trended segments of it (coloured). The linear fits used to obtain the de-trended fluxes are also shown in the same colour.
    Right panels: ACF on the {\it TESS} observed light curve (black, upper panel) and its linearly de-trended segments (coloured) shown in the corresponding left panels. }
    \label{acf}
\end{figure*}

In this sub-section we use this method to study the characteristic time-scales in the {\it TESS} light curve. 
The ACF run on the simple aperture photometry fluxes is shown in top-right panel of Fig.~\ref{acf}. 
We adopted a preliminary binning of the SAP\textunderscore FLUX of 30 min and then sampled the ACF in time separation  bins of 3 h. 
The only strong minimum in the ACF occurs at $\tau \sim 15.5$ d, which corresponds to the time distance between the flux minimum at $\rm JD \sim 2458849.5$ and the maximum at $\rm JD \sim 2458865$. The only prominent maximum besides that at the origin is at $\tau \sim 10.5$ d, the interval among various peaks or dips.
However, the results are strongly affected by the long-term trend of the {\it TESS} light curve, which can mask other signals. It is well known that blazars present different variability time-scales; if the goal is to investigate short time-scales, it is necessary to correct for the long-term flux oscillations before calculating the correlation function \citep[e.g.][]{smith1993}.

To perform this light-curve de-trending, we first subtracted best-fit linear trends to different parts of the light curve which show various decreasing or increasing baselines (see Fig.~\ref{acf}, top-left panel).
The dates of the connection points were chosen so that the de-trended light curve presents no jumps there. We then run the ACF on the de-trended light curve, which now reveals many more clear minima and maxima (top-right panel).
The higher peak beside the origin is at $\tau \sim 6.3 \rm \, d$, which is very similar to the minor peak at 6.5 d shown by the periodogram in Fig.~\ref{powert}. Other double-peaked, minor maxima are visible at 2.6--3.3 d, this too confirming the PSD results, and at 13.3--13.8 d.
The strongest minimum is at $\tau \sim 1.63 \rm \, d$, followed by $\tau \sim 4.4 \rm \, d$ and
$\tau \sim 15.5 \rm \, d$. Weaker signals appear at $\tau \sim 7.0 \rm \, d$, $\tau \sim 11.8 \rm \, d$, and $\tau \sim 12.5 \rm \, d$.
We note that these signals were already sketched in the ACF on the original {\it TESS} light curve, but they were overwhelmed by the long-term baseline trends.
In the same way, the imprint of further, shorter time-scales may be hindered by the prevalence of the major, day-long features in the de-trended light curve. To verify this point, we have visually identified segments of the {\it TESS} light curve, which are short enough to avoid big features, but covering at least 24 h each to explore time-scales shorter than 1 d. We show them in Fig.~\ref{acf}. After linear de-trending, we calculated the ACF with 10 min time resolution. The ACFs reveal a variety of sub-day time-scales, 
with minima around  $\tau \sim 0.2 \rm \, d$, 0.35--$0.40 \rm \, d$, and 0.5--$0.7 \rm \, d$.
Indeed, the {\it TESS} light curve results from the superposition of a great deal of flux oscillations of different duration and strength. 

The above {\it local} analysis on short segments of the {\it TESS} light curve revealed a wealth of {\it intra-day} variability time-scales,
suggesting that the process at the base of these flux changes can be described by a {\it stochastic} model. However, if we identified {\it characteristic} variability time-scales on the whole light curve, these would likely be due to persistent physical processes which can be treated by {\it deterministic} models. To verify this point, we develop a method for de-trending the {\it TESS} light curve for progressively shorter baseline trends to unveil progressively shorter characteristic time-scales.

\begin{figure*}
	\includegraphics[width=12cm]{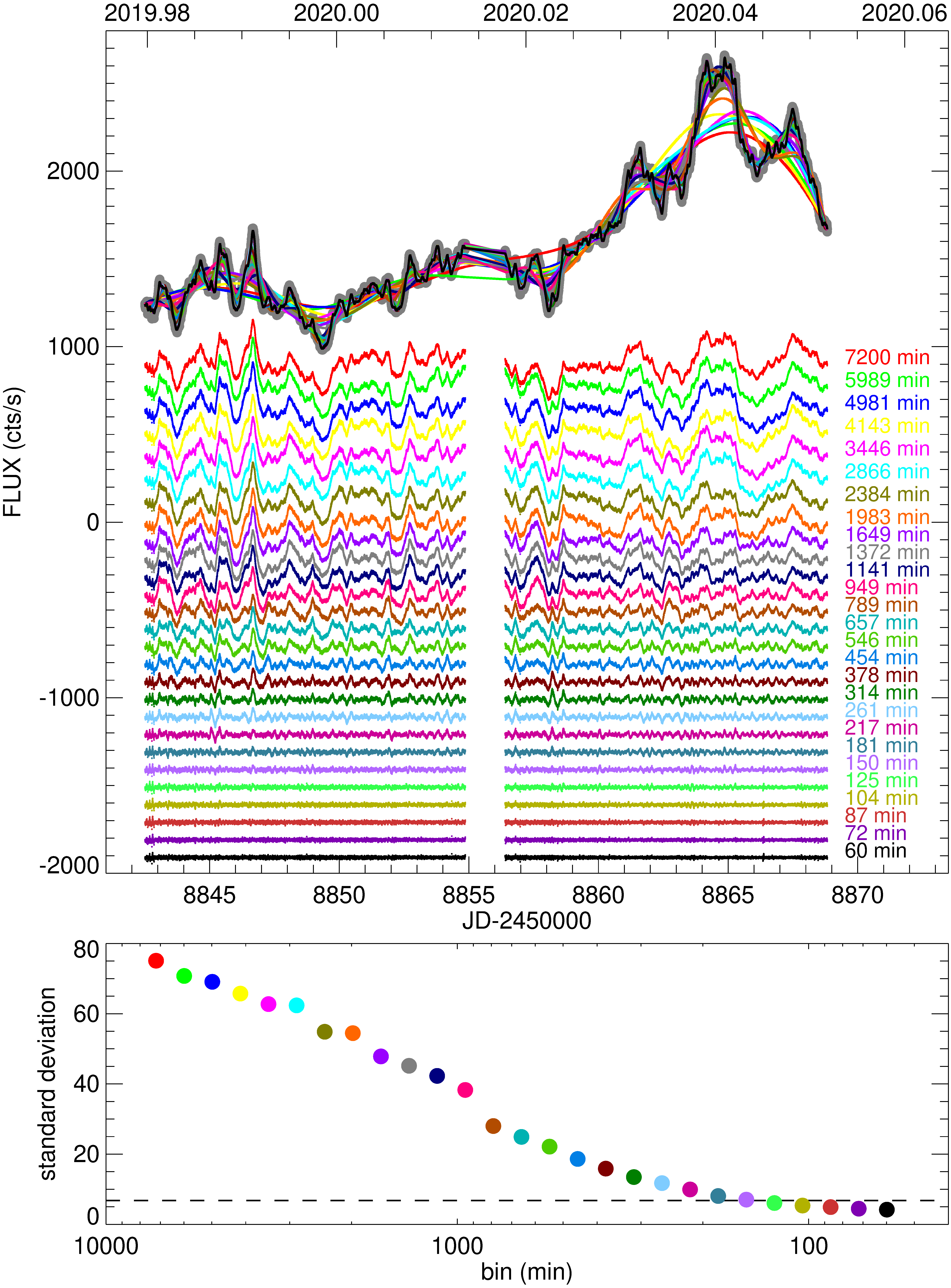}
    \caption{Upper panel: {\it TESS} observed light curve (grey dots) with superposed the cubic spline interpolations through its binned fluxes with 27 different time bins from 5 d (7200 min) to 1 h (60 min). The de-trended light curves obtained by dividing the {\it TESS} light curve by the splines are shown below (properly shifted to distinguish them). They show less and less variability as the time bin decreases.
    Bottom panel: standard deviation of the de-trended light curves shown in the upper panel. The horizontal dashed line indicates the mean error on {\it TESS} data.}
    \label{splines}
\end{figure*}

We use cubic spline interpolation with variable time binning as our de-trending tool.
We build 27 binned light curves from the observed {\it TESS} light curve. Each of them is obtained with a different binning time interval, ranging from 5 d to 1 h and equally spaced in logarithm.
The binned light curves are then interpolated with cubic splines\footnote{We note that cubic spline interpolations have been largely used in blazar variability studies to model the long-term trend \citep[e.g.][]{ghisellini1997,villata2002,raiteri2017_nature}.} (see Fig.~\ref{splines}).
De-trended light curves are finally obtained by dividing the observed {\it TESS} light curve by the splines and then multiplying for the same scale factor.
They are plotted in Fig.~\ref{splines} and show progressively smaller fluctuations as the splines enter more deeply inside the light curve ripples. The major flares clearly distinguishable in the 5 d (7200 min) de-trended light curve\footnote{Here and in the following, when we mention a given time de-trended light curve, we mean the light curve which is obtained by dividing the {\it TESS} light curve by the cubic spline interpolation through the {\it TESS} fluxes binned over that given time interval.} become less and less prominent; at time bins below 2400 min, their peaks are almost all at the same level, and below about 1200 min they have the same relevance as other events, which were minor flares in the observed light curve. For time bins of $\sim 660$ min, the signature of the major flares is completely lost and we can appreciate smaller and faster flux changes that were overwhelmed by more prominent flares in the observed light curve. As the binning interval continues to shorten, we are left with microvariations. Fig.~\ref{splines} shows that the standard deviation of the de-trended light curves with time bins below about 120 min is smaller than the mean error of the SAP fluxes. Yet, in the 60-min de-trended light curve we can still recognize well-defined flux variations that may still be of source intrinsic nature.
Indeed, in Section~\ref{secpsd} we found that instrumental white noise appears below a 14 min time-scale.

\begin{figure}
	\includegraphics[width=\columnwidth]{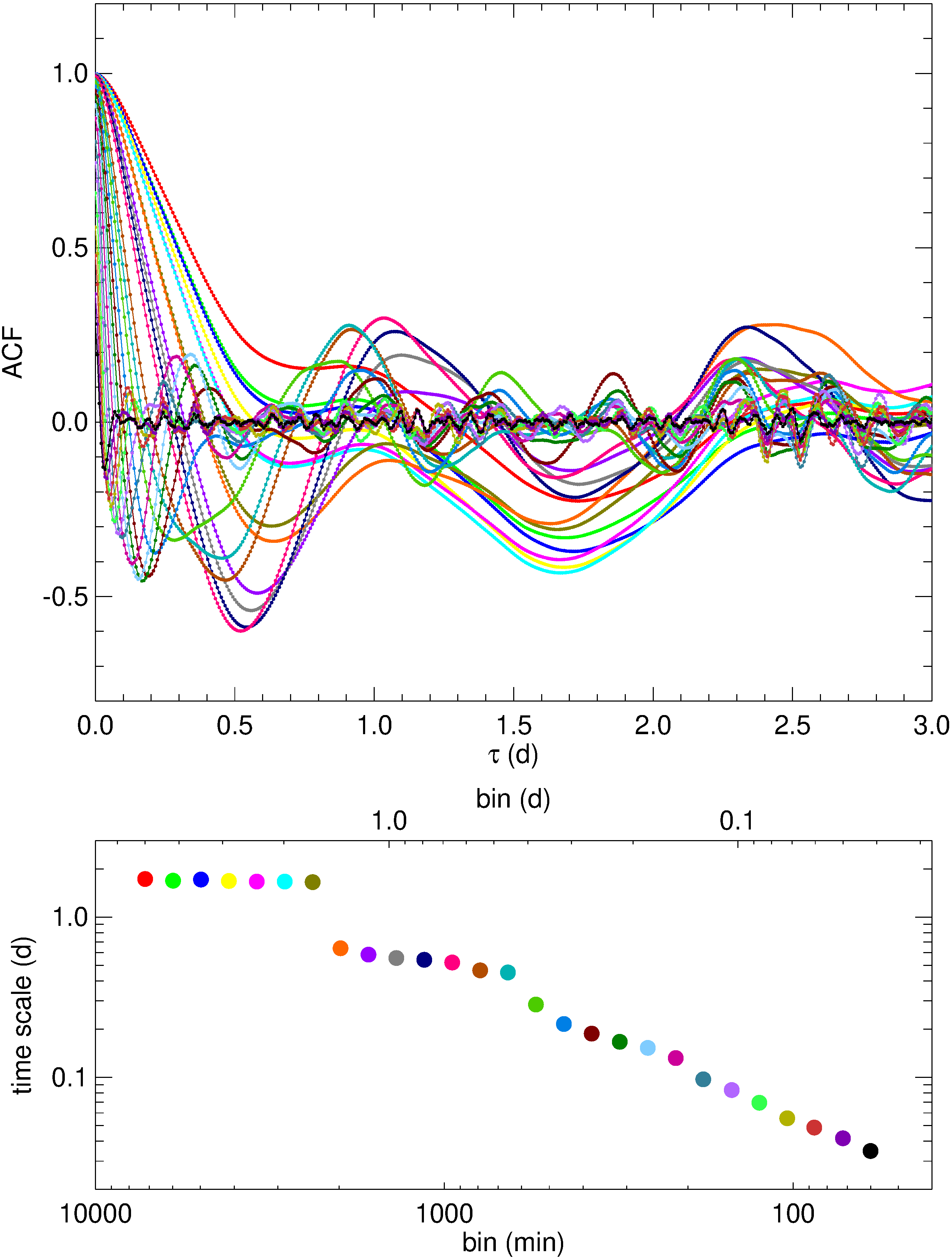}
    \caption{Upper panel: ACFs on the de-trended light curves shown in Fig.~\ref{splines}. Standard errors calculated according to \citet{edelson1988} and \citet{hufnagel1992} are within the lines thickness.
The most remarkable minima are found at time separations of about 0.17, 0.52, and 1.67 d.
Bottom panel: characteristic time-scales identified through minima in the ACFs of the de-trended light curves shown in the upper panel versus the time bin used to obtain the corresponding de-trending spline.}
    \label{tisca}
\end{figure}

The ACFs obtained from the de-trended light curves are presented in Fig.~\ref{tisca} (top panel). They have a time resolution of 10 min and explore time separations up to 3 d. We again obtain a variety of time-scales. The most remarkable minima, which are shared by several curves, are found at time separations of about 1.7, 0.5, and 0.2 d. 
The minimum time-scale decreases as the time bin used for the spline interpolation shortens, i.e.\ as the de-trending model follows the light curve in more and more detail.
We note that the ACFs with minima at $\sim 0.5$ d show maxima at about 1 d, suggesting repeating events on day time-scale, in agreement with the results of the PSD.

Fig.~\ref{tisca} (lower panel) shows the characteristic time-scale (i.e.\ the time separation at which the above ACFs present a minimum) versus the time bin which was used to obtain the spline representing the smoothed trend. 
This allows us to better define the most relevant time-scales.
Going from longer to shorter time bins, there are 7 de-trended light curves, corresponding to time bins from 7200 to 2384 min, whose deepest ACF minimum (for time separations less than 3 d) is found at $\tau \sim 1.67$ d, very close to that obtained with the previous linear de-trending. Then the principal minimum jumps at $\sim 0.52$ d. The ACF on the de-trended light curve corresponding to a time bin of 546 min (see Fig.~\ref{splines}) and to a time-scale of about 0.28 d, shows a somewhat transition behaviour. For shorter time bins the variability time-scale decreases almost linearly, reaching $\sim 0.035$ d for a 60 min time bin. As mentioned above, this is the case where the de-trended light curve still shows well-defined features, even if the standard deviation of the whole de-trended light curve is smaller than the mean error of the SAP fluxes. This almost linear decrease implies the existence of a nearly continuous set of time-scales from about 0.22 d to less than 1 h, among which the strongest signal corresponds to 0.17 d.

In conclusion, our method identifies three characteristic variability time-scales of 1.67, 0.52, and 0.17 d. These are revealed through the presence of common strong minima in the ACFs in the upper panel of Fig.~\ref{tisca} (and of steps in the lower panel). Indeed, such an aggregation means that these time-scales are true, intrinsic variability time-scales of the source, since they do not depend on the choice of a particular binning time interval, i.e.\ they are parameter-independent.

\begin{figure}
	\includegraphics[width=\columnwidth]{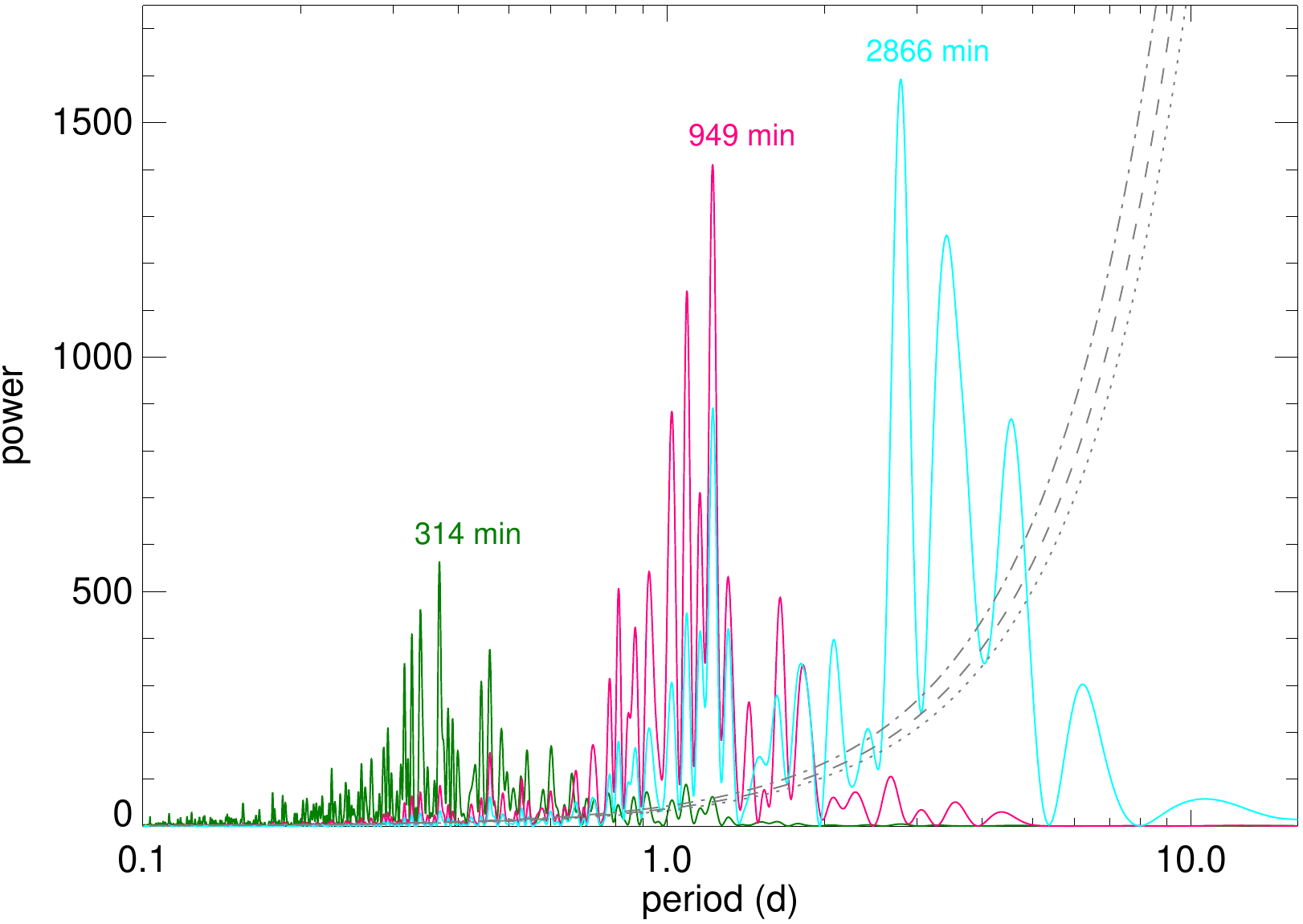}
    \caption{The periodogram of the three de-trended light curves in Fig.~\ref{splines} that show the deepest ACF minima in Fig.~\ref{tisca}.  These de-trended light curves were obtained by cubic spline interpolations through the {\it TESS} SAP fluxes binned in time intervals of 314 (dark green), 949 (magenta) and 2866 min (cyan). For comparison, the grey lines mark the significance levels of 95 (dotted), 99 (dashed) and 99.9 per cent (dot-dashed) reported in Fig.~\ref{powert}.}
    \label{power_dt} 
\end{figure}

It is now interesting to compare the results of the PSD and ACF analyses on the de-trended light curves.
Fig.~\ref{power_dt} shows the periodograms corresponding to the de-trended light curves obtained by using spline interpolations on the {\it TESS} fluxes binned over 2866, 949, and 314 min. These are the de-trended light curves that are characterized by the strongest ACF minima in Fig.~\ref{tisca}. By comparing these periodograms with that built with the {\it TESS} fluxes (Fig.~\ref{powert}), we see now a series of very high peaks, whose maxima are at 2.80, 1.22, and 0.368 d, respectively. Because peaks in the PSD correspond to ``periods", i.e.\ to time intervals between subsequent maxima (or minima), we must take their halves to compare them with the ACF characteristic time-scales, which instead give the interval between minima and maxima. We then obtain 1.40, 0.61, and 0.184 d, respectively, which are not far from the minima of the ACFs reported above.
We also notice that the periodogram corresponding to the 2866 min de-trended light curve also shows secondary maxima at the location of the main maxima in the PSD of the 949 and 314 min de-trended light curves, and in turn the PSD of the 949 min curve has a secondary maximum at the position of the principal peak of the 314 min curve. This underlines once again the fact that by decreasing the interpolation step we just emphasize signals at progressively shorter time intervals, which were already present in the PSD of the {\it TESS} SAP fluxes in Fig.~\ref{powert}.

\subsection{Structure function}
The structure function \citep[SF;][]{simonetti1985} is another method to analyse variability. It returns the square mean difference between flux densities as a function of their time separation $\tau$. Peaks in the SF indicate characteristic variability time-scales and should correspond to ACF dips.
Moreover, if $\rm SF \propto \tau^\beta$, the index $\beta$ is linked to the PSD spectral index $\alpha$ (see Section~\ref{secpsd}) by the following relationship: $\beta=\alpha-1$. Although the SF presents some caveats \citep{emmanoulopoulos2010}, we will use it here to see whether it can confirm the results obtained with the periodogram and auto-correlation function.

\begin{figure}
	\includegraphics[width=\columnwidth]{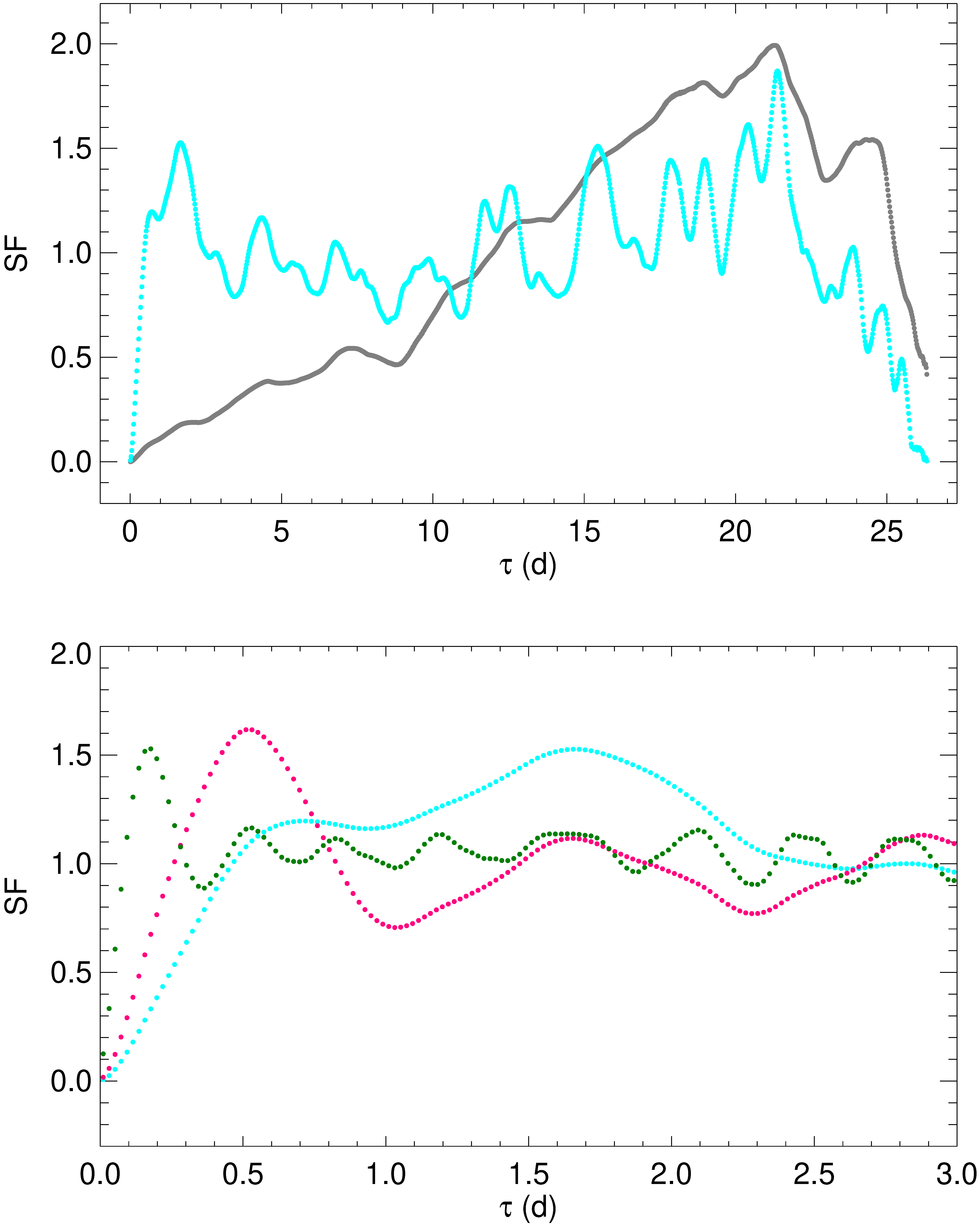}
    \caption{Top: Structure Function on the {\it TESS} data (grey) and on the 2866 min de-trended light curve shown in Fig.~\ref{splines} (cyan points). The SF on de-trended fluxes reveals many more characteristic time-scales, corresponding to the SF peaks.
    Bottom: SF on the 2866 (cyan), 949 (magenta) and 314 min (dark green) de-trended light curves for time lags shorter than 3 d. All SFs have been normalized to their average values.}
    \label{sf}
\end{figure}

The SF obtained from the {\it TESS} data is shown in the top panel of Fig.~\ref{sf}. It was calculated in bins of 30 min. 
The SF increases up to a time-scale of about 21 d, representing the maximum time-scale we can resolve. 
From the SF shape it is difficult to characterize variability, because there are no prominent maxima, especially at short time-scales. Therefore, we follow the same strategy adopted in Section~\ref{secacf} and work on de-trended light curves. 
We concentrate again on those which showed the strongest minima in their ACF below a time lag of 3 d. 
The SF on the 2866 min de-trended light curve is plotted in the same Fig.~\ref{sf} and reveals a lot of time-scales.
In particular, the first most remarkable peak is at $\tau=1.66 \rm \, d$, very close to the value inferred from the ACF. 
The bottom panel of Fig.~\ref{sf} shows again the SF on the 2866 min de-trended light curve together with those corresponding to 949 and 314 min binning of the original {\it TESS} fluxes. These SFs are zoomed for time lags shorter than 3 d and allow us to obtain time-scales of 0.51 and 0.18 d, in agreement with those already found when analysing the corresponding ACFs.

\begin{figure}
	\includegraphics[width=\columnwidth]{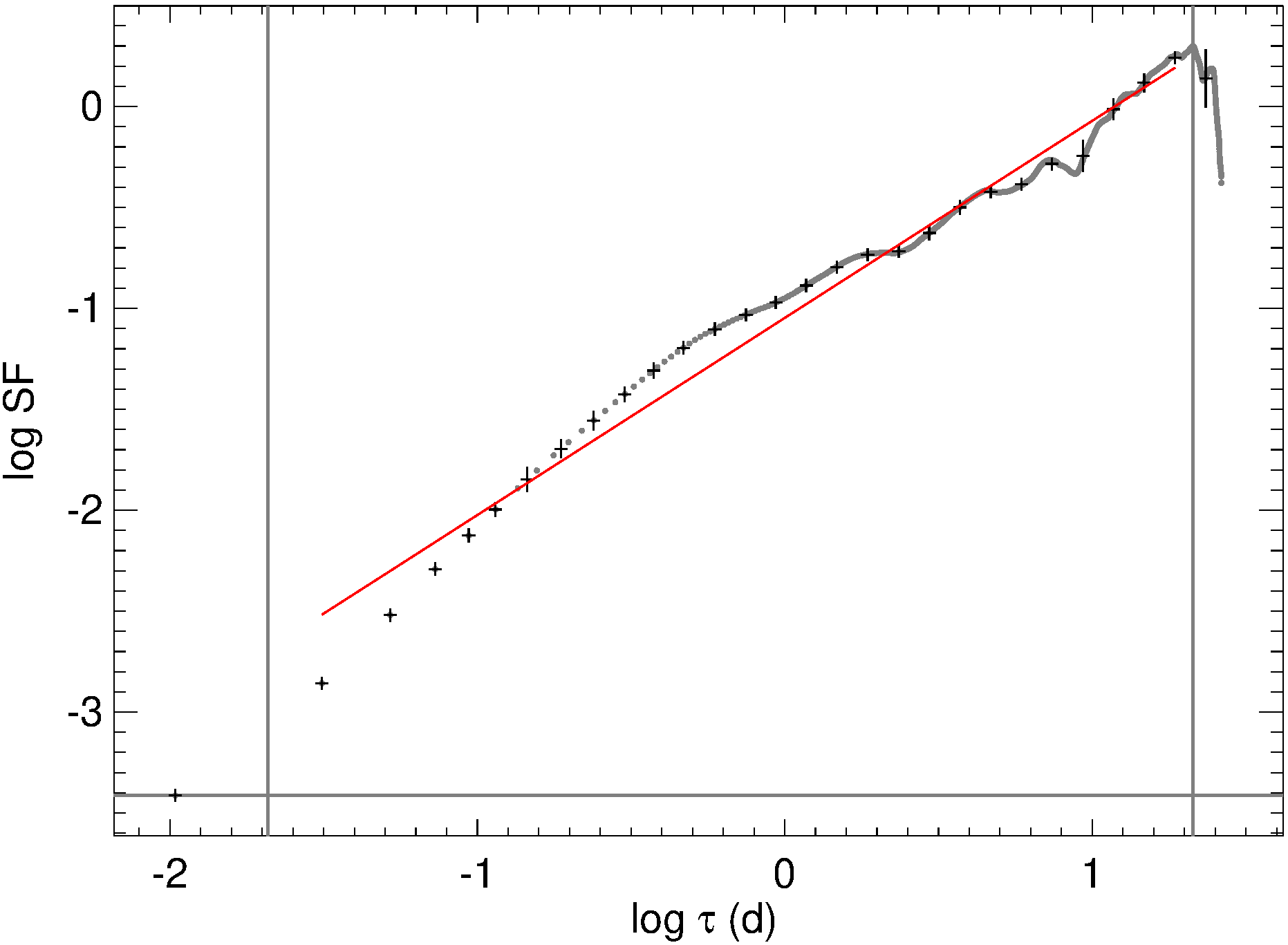}
    \caption{Structure Function on the {\it TESS} data in the log-log plot. To avoid oversampling at larger $\tau$ values, the SF has been binned (black plus signs) before fitting it with a straight (red) line between 14 minutes and 21.3 d (see text for details). The slope is $\beta=0.98 \pm 0.03$. }
    \label{sf_log}
\end{figure}

In Fig.~\ref{sf_log} the SF is plotted in logarithmic scale to measure its power-law index $\beta$ and thus investigate its relationship with the PSD.
As in the case of the PSD,
we binned the SF before fitting it with a power law to avoid now oversampling of the largest $\tau$ bins. 
Then we set $\tau$ limits for the fit: the lower is fixed at 30 min, the bin used to build the SF,
and the higher at 21.3 d, corresponding to the SF peak. The best-fit spectral slope is $\beta=0.98 \pm 0.03$, which matches the expected relationship with the PSD spectral index.

\section{Chromatic versus achromatic variations}
\label{seccro}
We now apply the de-trending method to investigate in more detail the chromatism of the flux variations on different time-scales.
In order to perform de-trending of the multiband WEBT data, we use the much better sampled {\it TESS} light curve and ``adapt" it to give a good description of the WEBT light curves. This is obtained by amplifying the {\it TESS} fluxes according to $F_{\rm amp}=F^\zeta$.  The amplification power $\zeta$ is set to 1.05 for the $R$ band and 1.15 for the $B$ band to match the amplitude of the corresponding flux density variations. The amplified light curves are then binned and interpolated with cubic splines. These splines are used to de-trend the flux densities: $F'_{R,B}=F_{R,B}/f_{R,B} \times {\rm min}(f_{R,B})$, with $f_{R,B}$ representing the spline fluxes at the acquisition times of the WEBT data. De-trended $R$ and $B$ magnitudes are finally derived from flux densities and colour indices are obtained with the same prescriptions adopted in Section~\ref{secwebt}.

\begin{figure}
	\includegraphics[width=\columnwidth]{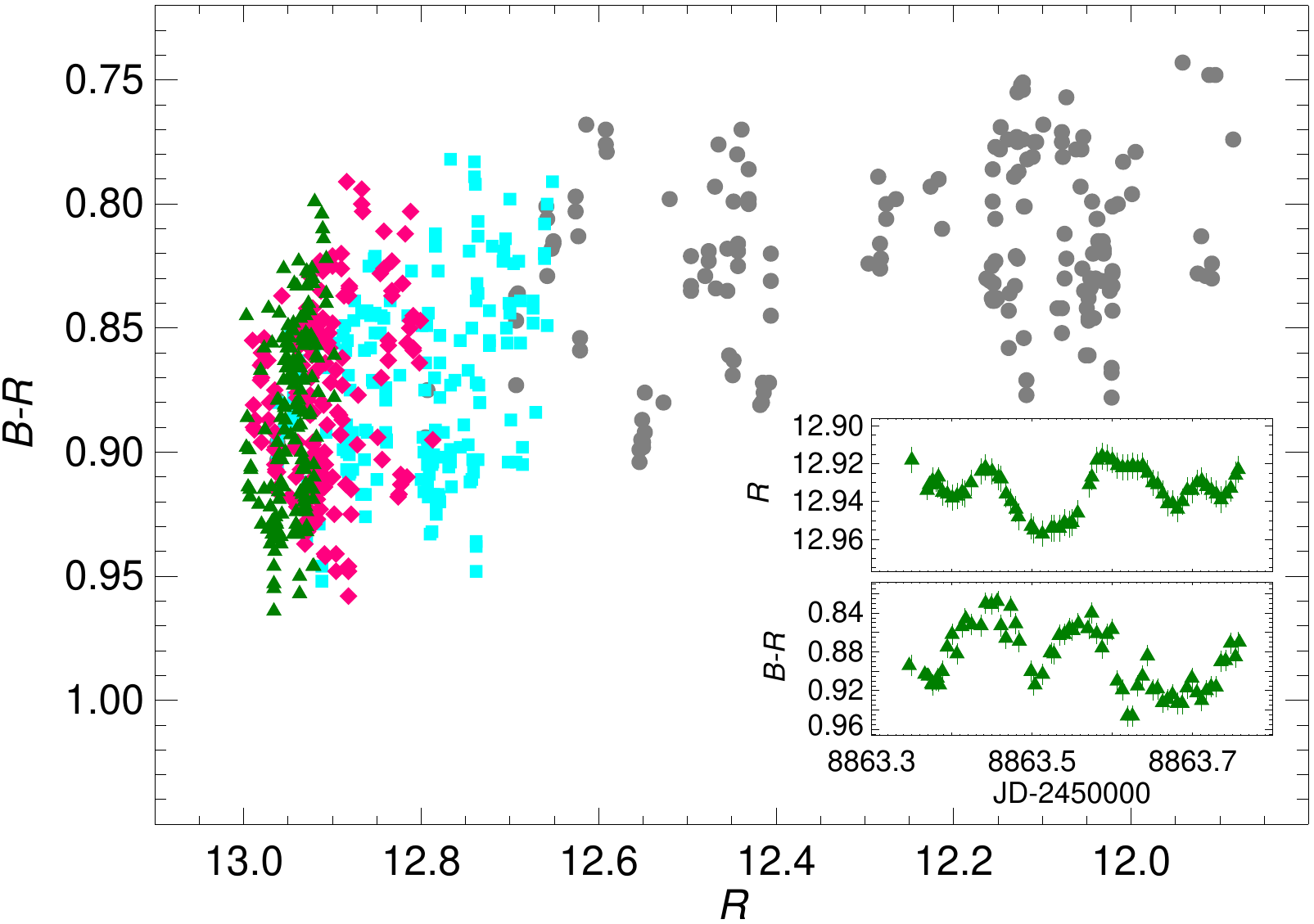}
    \caption{The colour index $B-R$ versus brightness obtained from the original WEBT light curves (grey dots) and in the cases of de-trended light curves. The cyan squares refer to de-trending with 2866 min time bins, magenta diamonds to that with 949 min time bins, and dark green triangles to that with time bins of 314 min. As the time bin shortens, the brightness range decreases, but the colour index range remains the same, confirming the strong chromatism of the short-term flux changes. In the inset we show the $R$-band light curve and $B-R$ indices versus time corresponding to the 314 min de-trended WEBT light curves on $\rm JD-2450000=8863$.}
    \label{bmr}
\end{figure}

Fig.~\ref{bmr} compares the colour indices inferred from the original WEBT data and already shown in Fig.~\ref{colori} to those calculated after de-trending the WEBT light curves as outlined above. The usual three cases corresponding to time intervals of 2866, 949, and 314 min for the {\it TESS} light curve binning and its spline interpolation are displayed. As the time bin shortens, the range of magnitude variation decreases, from about 1 mag down to 0.1 mag in the 314 min case. Instead, the range of values spanned by the colour indices remains about the same, i.e.\ $\sim 0.16$ mag. This is another confirmation that fast flares are strongly chromatic with respect to the flux changes on longer time-scales. The figure also shows the intra-night colour indices obtained from the 314 min de-trended WEBT light curves on $\rm JD-2450000=8863$ (see right panels of Fig.~\ref{colori} for a comparison to the situation before de-trending). The $R$-band magnitude range is 0.04, while the $B-R$ range is 0.12, yielding a $B-R$ versus $R$ slope of 3. This can be compared with the quasi-achromatism of the long-term oscillations, with slope 0.035, as estimated in Section~\ref{webt}.

\section{Discussion and Conclusions}
\label{seclast}

We have presented the 2-min cadence optical light curve of the BL Lac object S5~0716+714 obtained by the {\it TESS} satellite during cycle 2, from 2019 December 24 to 2020 January 21.
These data have been complemented by multiband optical monitoring carried out by the WEBT Collaboration and by ultraviolet and X-ray observations by the {\it  Neil Gehrels Swift Observatory}. While the {\it TESS} data allowed us to analyse the source variability down to the shortest time-scales, those collected by the WEBT and {\it Swift} provided information on the spectral variability. 
A simple power-law fit to the PSD and SF gives a slope of $\sim 2$ and $\sim 1$, respectively, suggesting that the variability process is in agreement with random walk/red noise, with more power at longer periods. 
However, both the PSD and SF shapes suggest that a more complex scenario must be envisaged to explain the source flux variations, as already found by other authors working with {\it Kepler} data \citep{edelson2013,sasada2017,goyal2018}. 
Interestingly, random walk noise was found by \citet{hughes1992} when analysing the radio light curves (at 4.8, 8.0 and 14.5 GHz) of a sample of 51 BL Lacs and FSRQs monitored for more than 10 yr at the University of Michigan Radio Astronomy Observatory (UMRAO).
The fact that optical and radio PSD share the same properties is expected, because the emission in both these bands is due to a synchrotron process, though likely coming from separate regions in jet. The observed smoother radio variability, 
the commonly detected delay of the radio flares after the optical ones, together with synchrotron self-absorption being stronger at radio wavelengths, make the observed radio emission likely come from more external jet zones than the optical radiation. 

From the colour indices we have found that the long-term optical variability of S5~0716+714 is characterized by a mild chromatism. This is compatible with the major flux variations be due to changes of the Doppler factor. In contrast, fast flux changes are strongly chromatic and are likely of intrinsic and energetic nature. The X-ray spectrum is always soft, implying a dominant contribution by the synchrotron component. However, the smallest spectral index belongs to the faintest state, in agreement with the contribution of the inverse-Compton becoming more appreciable when the brightness decreases. We detected significant fast variations also in the {\it Swift} data: up to $\ga 0.2$ mag in the ultraviolet in 0.26 d, and more than a factor 2 in the X-ray flux in 0.27 d.

Time series analysis techniques on the {\it TESS} data can reveal the shortest characteristic time-scales of variability. However, this can be achieved only after the removal of the overwhelming signature of the larger flux changes.
Light curve de-trending is a necessary step to unveil the time-scales of the fastest flux variations, but the choice of the de-trending function can affect the results of the analysis. Therefore, we have proposed a method 
that performs de-trending by using cubic splines interpolations through the {\it TESS} light curve binned in progressively shorter time intervals. As the time bin decreases, the interpolation follows in more and more detail the meanderings of the light curve, and the de-trending process removes larger variations emphasizing the smaller ones.
The de-trended light curves then have been analysed by means of the periodogram, the auto-correlation function, and the structure function.
We have found that the shortest characteristic time-scales (typical time intervals between a flux maximum and a minimum or viceversa), are about 1.7, 0.5 and 0.2 d long. Table~\ref{sum} summarizes the results obtained with the various methods.
\begin{table}
	\centering
	\caption{Summary of the shortest characteristic time-scales obtained from the {\it TESS} SAP fluxes by means of the power spectral density (PSD), auto-correlation function (ACF), and structure function (SF). In the case of the PSD, we indicated half of the periods that appear marginally significant in the periodogram. ``Bin" represents the time binning interval used to obtain the de-trended light curve that revealed the signal in the strongest way.}
	\label{sum}
	\begin{tabular}{rrrr} 
		\hline
		PSD & ACF & SF & Bin \\
                (d) & (d) & (d) & (min) \\
		\hline
                0.18 & 0.17 & 0.18 & 314\\
		0.61 & 0.52 & 0.51 & 949\\
		1.40 & 1.67 & 1.66 & 2866\\
		\hline
	\end{tabular}
\end{table}

Below about 0.2 d, the source flux shows variability on a continuous variety of time-scales, suggesting a process that can be described by a stochastic model. 
Causality arguments imply that the shortest variability time-scales set an upper limit to the dimension $D$ of the emitting region, 
$D \le c \Delta t_{\rm obs} \delta / (1+z)$, where $z$ is the redshift. For a source at $z \sim 0.3$ like S5~0716+714, and with a value of $\delta \sim 10$ \citep{savolainen2010,kravchenko2020}, a 0.2 d time-scale means that the region must be smaller than about $10^{-3} \rm \, pc$. 
This would imply a jet sub-structure, if we assume that the jet dimension where the radiation is emitted is that predicted by blazar SED modelling, i.e.\ typically of the order of hundredths of pc in case it is inside the broad line region, and of the order of tenths of pc if it is outside \citep[see e.g.][and references therein]{acciari2018}. 
Therefore, the strongly chromatic flux variations on time-scales shorter than about 0.2 d must have an intrinsic nature and are likely linked to sub-structures of the jet. On  these time-scales we may see the stochastic footprint of injection/acceleration of particles in turbulent jet regions \citep{marscher2014,pollack2016}. 

On longer time-scales, where flux variability is quasi-achromatic, we likely see the effect of variations of the Doppler factor. As in previous works, we favour a scenario in which $\delta$ changes are caused by variations of the viewing angle of the emitting jet regions. Time-scales as short as 0.5 d may still imply sub-structures of the jet, which we may imagine as twisted filaments. In this view, the variability time-scales would correspond to the time a filament takes to align close to the line of sight.
The de-trending procedure applied in this paper is consistent with correcting for variable beaming in case the flux variations are due to changes in the Doppler factor. 
Indeed, assuming that $\delta$ changes in time, the ratio between two observed flux densities will depend on some power of the ratio between their Doppler factors\footnote{The observed flux density $F_\nu \propto \delta^{n+\alpha}$, where $\alpha$ is the spectral index and $n=2$ for a continuous jet, while $n=3$ for a discrete point-like component \citep{urry1995}.}. If we represent the observed baseline flux with a cubic spline interpolation through the binned light curve, then by dividing the fluxes by the spline and normalizing to some minimum flux, we obtain a Doppler-corrected light curve, i.e.\ we remove that part of variability caused by variations of the Doppler factor. 

In previous works we already found that an interpretation of the variability in terms of variations of the Doppler factor can successfully be applied to the long-term behaviour of many blazars 
\citep[e.g.][]{villata1998b,villata1999,villata2004,ostorero2006,villata2009a,raiteri2010,raiteri2011,raiteri2012,raiteri2013,raiteri2015}, most remarkably to the multifrequency variability of CTA~102 \citep{raiteri2017_nature}. In all these cases, we made the further assumption that the changes of the Doppler factor have a geometric explanation, i.e.\ that they are due to a variation of the orientation of the jet emitting regions. 
Here for the first time we can quantify what is the shortest variability time-scale due to geometrical effects. We found that in S5~0716+714 it corresponds to about half a day, i.e.\ $\sim 4$ d in the rest frame, if $\delta \sim 10$.
The existence of various geometric time-scales can be explained if we think of a twisted jet that is composed of many filaments winding up around the jet axis like strands in a rope. Apart from the case where the viewing angle of the jet axis is null, each filament would have its own orientation with respect to the observer and thus its own changing Doppler factor. 
Moreover, each strand may be made up of thinner threads, giving rise to different variability time-scales.

A geometric interpretation of the S5~0716+714 variability was also given by \citet{kravchenko2020}. They analysed space-VLBI and VLBA observations, which revealed that the jet has a curved structure. These authors favoured a scenario where radio variability is due to Doppler factor changes caused by changes of the viewing angle.

A final remark is due on the issue whether the shape of flux variations is symmetric or not. According to the scenario traced above, we would expect that the intrinsic and chromatic flux changes on time-scales below about 0.2 d be asymmetric.
In contrast, the mildly achromatic variations on longer time-scales caused by orientation changes should be symmetric.
The point is hard to investigate, because most events are blended, i.e.\ they seem the result of the superposition of multiple events. 
Moreover, we could speculate on whether variations of the Doppler factor of geometrical origin always produce flux changes with symmetric profiles. This is certainly true when the viewing angle changes in a ``regular" way, as in a helical rotating jet with constant rotation velocity. But if the change in the orientation of the emitting jet region occurs in an irregular way, 
as in the case of different velocities in approaching/leaving the line of sight, then the resulting flux variation profile could be asymmetric.

\section*{Acknowledgements}
The Astronomical Observatory of the University of Siena thanks the friend amateur astronomers Massimo Conti and Claudio Vallerani for their invaluable and unceasing contribution, essential for the performance of the observatory.
This paper includes data collected by the {\it TESS} mission. Funding for the {\it TESS} mission is provided by the NASA Explorer Program.
This research has made use of NASA's Astrophysics Data System and of the NASA/IPAC Extragalactic Database (NED), which is operated by the Jet Propulsion Laboratory, California Institute of Technology,
under contract with the National Aeronautics and Space Administration.
Based on observations made with the Nordic Optical Telescope, operated by the Nordic Optical Telescope Scientific Association at the Observatorio del Roque de los Muchachos, La Palma, Spain, of the Instituto de Astrofisica de Canarias.
Based (partly) on data obtained with the STELLA robotic telescopes in Tenerife, an AIP facility jointly operated by AIP and IAC.
This work is partly based upon observations carried out at the Observatorio Astron\'{o}mico Nacional on the Sierra San Pedro Martir (OAN-SPM), Baja California, Mexico. 
This research was partially supported by the Bulgarian National Science Fund of the Ministry of Education and Science under grants DN
18-13/2017, KP-06-H28/3 and KP-06-PN38/4. 
K.~M. acknowledges JSPS KAKENHI grant no. JP19K03930.
S.~O.~K. acknowledges financial support by Shota Rustaveli National Science Foundation of Georgia under contract PHDF-18-354.
E.~B. acknowledges support from DGAPA-PAPIIT GRANT IN113320.
G.~B. acknowledges financial support from the State Agency for Research of the Spanish MCIU through the
``Center of Excellence Severo Ochoa" award to the Instituto de Astrof\'{i}sica de Andaluc\'{i}a (SEV-2017-0709).

\section*{Data availability}
The data collected by the WEBT collaboration are stored in the WEBT archive at the Osservatorio Astrofisico di Torino - INAF (http://www.oato.inaf.it/blazars/webt/); for questions regarding their availability, please contact the WEBT President Massimo Villata ({\tt massimo.villata@inaf.it}).




\bibliographystyle{mnras}
\bibliography{pate} 


\section*{Affiliations}
{\it
$^{ 1}$INAF, Osservatorio Astrofisico di Torino, Via Osservatorio 20, I-10025 Pino Torinese, Italy                                                           \\
$^{ 2}$EPT Observatories, Tijarafe, E-38780 La Palma, Spain                                                                                                  \\
$^{ 3}$INAF, TNG Fundaci\'on Galileo Galilei, E-38712 La Palma, Spain                                                                                        \\
$^{ 4}$Instituto de Astronom\'{i}a, Universidad Nacional Aut\'{o}noma de M\'{e}xico, Cd.\ Universitaria, CDMX, Mexico                                        \\
$^{ 5}$Abastumani Observatory, Mt.\ Kanobili, 0301 Abastumani, Georgia                                                                                       \\
$^{ 6}$Samtskhe-Javakheti State University, 92 Shota Rustaveli St.\ Akhaltsikhe, Georgia                                                                     \\
$^{ 7}$Landessternwarte, Zentrum f\"ur Astronomie der Universit\"{a}t Heidelberg, K\"{o}nigstuhl 12, 69117 Heidelberg, Germany                               \\
$^{ 8}$Aryabhatta Research Institute of Observational Sciences (ARIES), Manora Peak, Nainital - 263 001, India                                               \\
$^{ 9}$Ulugh Beg Astronomical Institute, Maidanak Observatory, Tashkent, 100052, Uzbekistan                                                                  \\
$^{10}$INAF, Istituto di Radioastronomia, Via Gobetti 101, I-40129 Bologna, Italy                                                                            \\
$^{11}$Pulkovo Observatory, 196140 St.\ Petersburg, Russia                                                                                                   \\
$^{12}$Nordic Optical Telescope, Apartado 474, E-38700 Santa Cruz de La Palma, Spain                                                                         \\
$^{13}$Instituto de Astrof\'{i}sica de Canarias, E-38200, La Laguna, Tenerife, Spain                                                                         \\
$^{14}$Universidad de La Laguna, Departamento de Astrofisica, E38206, La Laguna, Tenerife, Spain                                                             \\
$^{15}$Crimean Astrophysical Observatory RAS, P/O Nauchny, 298409, Russia                                                                                    \\
$^{16}$University of Siena, Department of Physical Sciences, Earth and Environment, Astronomical Observatory, Via Roma 56, I-53100 Siena, Italy              \\
$^{17}$Graduate Institute of Astronomy, National Central University, 300 Jhongda Road, Zhongli, Taoyuan 32001, Taiwan                                        \\
$^{18}$Osservatorio Astronomico Sirio, I-70013 Castellana Grotte, Italy                                                                                      \\
$^{19}$Instituto de Astronom\'{i}a, Universidad Nacional Aut\'{o}noma de M\'{e}xico, Ensenada, Baja California, Mexico                                       \\
$^{20}$Engelhardt Astronomical Observatory, Kazan Federal University, Tatarstan, Russia                                                                      \\
$^{21}$Facultad de Ciencias, Universidad Aut\'{o}noma de Baja California, Campus El Sauzal, Ensenada B.C., Mexico                                            \\
$^{22}$Astronomical Institute, Osaka Kyoiku University, Osaka 582-8582, Japan                                                                                \\
$^{23}$Instituto Nacional de Astrof\'{i}sica, \'{O}ptica y Electr\'{o}nica, Tonantzintla, Puebla, Mexico                                                     \\
$^{24}$Department of Physics and Mathematics, Aoyama Gakuin University, 5-10-1 Fuchinobe, Chuo-ku, Sagamihara, Kanagawa 252-5258, Japan                      \\
$^{25}$Astronomical Institute, St.\ Petersburg State University, 198504 St.\ Petersburg, Russia                                                              \\
$^{26}$Institute of Astronomy and National Astronomical, Observatory, Bulgarian Academy of Sciences, Sofia, Bulgaria                                         \\
}






\bsp	
\label{lastpage}
\end{document}